\input harvmac.tex
\let\includefigures=\iftrue
\newfam\black
\includefigures
\input epsf
\def\figin{\epsfcheck\figin}\def\figins{\epsfcheck\figins}
\def\epsfcheck{\ifx\epsfbox\UnDeFiNeD
\message{(NO epsf.tex, FIGURES WILL BE IGNORED)}
\gdef\figin##1{\vskip2in}\gdef\figins##1{\hskip.5in}
\else\message{(FIGURES WILL BE INCLUDED)}%
\gdef\figin##1{##1}\gdef\figins##1{##1}\fi}
\def\DefWarn#1{}
\def\figinsert{\goodbreak\midinsert}
\def\ifig#1#2#3{\DefWarn#1\xdef#1{fig.~\the\figno}
\writedef{#1\leftbracket fig.\noexpand~\the\figno}%
\figinsert\figin{\centerline{#3}}\medskip\centerline{\vbox{\baselineskip12pt
\advance\hsize by -1truein\noindent\footnotefont{\bf Fig.~\the\figno:} #2}}
\bigskip\endinsert\global\advance\figno by1}
\else
\def\ifig#1#2#3{\xdef#1{fig.~\the\figno}
\writedef{#1\leftbracket fig.\noexpand~\the\figno}%
\global\advance\figno by1}
\fi

\def\ka{K\"ahler}

\def\p{{\bf{P}}}
\def\f{{\bf{F}}}
\def\s{{\bar{S}}}

\def\np#1#2#3{{Nucl. Phys. {\bf B{#1}} ({#2}) {#3}}}
\def\phase#1{{\widetilde{\rm {#1}}}}
\Title{\vbox{\baselineskip12pt\hbox{hep-th/9807170}
\hbox{RU-98-27}}}
{\vbox{
\centerline{Calabi-Yau Spaces and Five Dimensional Field Theories}
\vskip 10pt
\centerline{with Exceptional Gauge Symmetry}}}
\vskip 10pt
\centerline{Duiliu-Emanuel Diaconescu
and Rami Entin}
\medskip
\centerline{\it Department of Physics and Astronomy}
\centerline{\it Rutgers University }
\centerline{\it Piscataway, NJ 08855--0849}
\centerline{\tt duiliu, rami@physics.rutgers.edu}
\medskip
\bigskip
\noindent

Five dimensional field theories with exceptional gauge groups are 
engineered from degenerations of Calabi-Yau threefolds. The 
structure of the Coulomb branch is analyzed in terms of relative 
\ka\ cones. For low number of flavors, the geometric construction 
leads to new five dimensional fixed points. 

\Date{July 1998}

\newsec{Introduction and Summary}

Five dimensional gauge theories with ${\cal{N}}=1$ supersymmetry are 
in general not renormalizable and therefore should be viewed as
theories with a cut-off. An exception to this occurs if a theory is 
defined by an interacting UV fixed point of the renormalization
group. Concrete examples with one dimensional Coulomb branch 
have been discovered for the first time in 
\ref\SEI{N. Seiberg,
``Five-Dimensional SUSY Field Theories, Nontrivial Fixed Points and
String Dynamics'', \np{483}{1997}{229},
hep-th/9608111.} by studying D4-branes near orientifold fixed planes.
The same theories are associated to del Pezzo contractions in
Calabi-Yau threefolds in
\nref\MS{N. Seiberg and
D. R. Morrison, ``Extremal Transitions and Five-Dimensional
Supersymmetric Field Theories'', \np{483}{1997}{229},
hep-th/9609070.}%
\nref\DKV{M. R. Douglas,
S. Katz and C. Vafa, ``Small Instantons, del Pezzo Surfaces and Type
I$^\prime$ Theory'', \np{497}{1997}{155},
hep-th/9609071}%
\refs{\MS,\DKV}. 

Theories with higher rank gauge groups and their corresponding
Calabi-Yau degenerations were considered in 
\nref\IMS{K. Intriligator,
D.R. Morrison and N. Seiberg, ``Five-Dimensional Supersymmetric Gauge
Theories and Degenerations of Calabi-Yau Spaces'', \np{497}{1997}{56},
hep-th/9702198.}%
\nref\Nek{N. Nekrasov, ``Five Dimensional Gauge Theories and
Relativistic Integrable Systems'',  hep-th/9609219.}%
\refs{\IMS,\Nek}.
The exact quantum prepotential for
general gauge groups and matter content was determined, and complete
list of possible fixed points with a gauge theory origin was given. 
A necessary condition for the existence of a fixed point is that the
metric on the Coulomb branch will be non-negative which
gives an upper bound on the allowed number of matter representations. 
Alternatively, one can use Higgsing arguments to limit the number of
flavors in the higher rank theory if the bound in the lower rank
theory is known. The precise agreement found
between the gauge theory calculation of the prepotential and its 
geometric counterpart allows one to address the issue of existence of
fixed points directly in the geometry. In some cases it is possible to
determine that a necessary condition is also sufficient. This was done
by \IMS\ for certain classical groups with restricted matter representations.
An alternative approach based on brane constructions appeared in 
\nref\AH{O. Aharony and A. Hanany, ``Branes, Superpotentials and
Superconformal Fixed Points'', \np{504}{1997}{239},
hep-th/9704170.}%
\nref\AHK{O. Aharony, A. Hanany and B. Kol, ``Webs of (p,q) 5-branes, 
Five Dimensional Field Theories and Grid Diagrams'', hep-th/9710116.}
\refs{\AH,\AHK}. As shown in \ref\LV{N.C. Leung and C. Vafa,
``Branes and Toric Geometry'', hep-th/9711013.} all the theories that
arise this way can also be realized by compactifying M-theory on
torically degenerated Calabi-Yau spaces.

The subject of the present paper is exceptional gauge group theories 
with matter content in the smallest allowed
representations. As in \IMS\ we find a complete agreement between the 
gauge theory prepotential and the one computed from triple
intersections of the corresponding Calabi-Yau degeneration. Our main
result concerns the existence of fixed points. We perform in detail
the geometric analysis and show that some of the necessary conditions
found in \IMS\ are also sufficient. Concretely, we show the
existence of the following fixed point theories
\eqn\fpoint{\eqalign{
G_2&\qquad n_7\leq4\cr
F_4&\qquad n_{26}\leq3\cr
E_8&\qquad n_{248}=0\cr
E_6&\qquad n_{27}\leq3\cr
E_7&\qquad n_{{1\over{2}}56}\leq5}.}
The Coulomb branches of the $E_6$ and $E_7$ theories turn out to have an
interesting phase structure already in the presence of massless
matter. Finally, as expected by the absence of a certain global
anomaly in the $E_7$ gauge theory, we find that theories with an odd
number of half ${\bf 56}$ hypermultiplets do occur.

The plan of this paper is as follows. In the rest of this section we
give a short summary of the properties of five dimensional gauge
theories we will need and briefly review the geometry-gauge theory 
correspondence. The relatively simple cases of $G_2$, $F_4$ and 
$E_8$ gauge theories are discussed in section 2. Sections 3 and 4 
are devoted to the $E_6$ and $E_7$ theories respectively. Appendix A
contains a summary of the relevant facts about ruled surfaces. The
canonical resolutions of the $E_6$ and $E_7$ singularities which are
the starting points of constructing the corresponding Calabi-Yau 
degenerations are given in appendices B and C.

\subsec{Five dimensional gauge theories}

Following \refs{\SEI,\MS,\IMS}, we recall some relevant features of
five dimensional gauge theories. 
Along the Coulomb branch, the effective low energy description of a
theory with an exceptional gauge group $G$ of rank $r$ is that of 
an Abelian gauge theory with $r$ $U(1)$ gauge fields. It is determined
in terms of the
prepotential ${\cal F}(\phi^i)$ where $\phi^i$ are the scalar partner
of the $r$ $U(1)$ vectors. Five dimensional gauge invariance
constrains $\del_i\del_j\del_k{\cal F}$ to be integral which restricts
the local form of the prepotential to be at most cubic in $\phi^i$. 
Therefore the exact quantum prepotential is determined already at 
one-loop. For massless matter in representations ${\bf r}_f$ of $G$, 
it is given by\foot{We consider exceptional gauge groups whose $d$ 
symbol vanishes, making a classical Chern-Simons term impossible.}
\eqn\exact{{\cal
F}={1\over{2}}m_0h_{ij}\phi^i\phi^j+{1\over{12}}
\left(\sum_{{\bf r}\in{\bf R}}
|{\bf r}\cdot\phi|^3-\sum_f\sum_{{\bf w}\in {\bf W}_f}|{\bf w}
\cdot\phi|^3\right),}
where $h_{ij}$ is the trace ${\rm Tr}\,T_iT_j$ of the Cartan
generators, and ${\bf R}$ and ${\bf W}_f$ are the weight systems of
the adjoint and matter representations.
The first two terms are present at the classical level. The last two 
are generated quantum mechanically by integrating out massive gauge 
bosons and charged matter which contribute with opposite signs to the 
prepotential. 

The signs of the ${\bf r}\cdot\phi$ terms are identical for all ${\bf
r}\in{\bf R}$ in any 
given Weyl chamber. Choosing a definite sign for the first quantum 
term in \exact\ gives an expression which
is valid throughout the Weyl chamber. A non-trivial phase structure
emerges when ${\bf w}\cdot\phi$ vanishes along certain codimension 
one boundaries inside the Weyl chamber, creating a wedge structure.
Since these terms enter the prepotential with an absolute value
there are different prepotentials in each of the sub-wedges. The
metric on the Coulomb branch is still continuous since it is
determined by the Hessian $g_{ij}=\del_i\del_j{\cal{F}}$.

As mentioned earlier, a necessary condition for the existence of fixed
point is that the metric should be non-negative on the entire Coulomb
branch.  Only then it is possible to have a sensible quantum theory on
the entire moduli space, since a breakdown of the metric is a sign of
non-renormalizability. This condition is equivalently expressed as the
requirement that $\cal{F}$ be a convex function on the entire Weyl
chamber. The gauge bosons term is obviously convex since it enters 
the prepotential with a positive sign. The negative sign of the matter
contribution makes that term concave. Thus, a necessary condition for
the existence of fixed points will be in the form of an upper bound on
the number of flavors.

\subsec{Gauge theory from geometry}

Here we review the basic ideas of geometric engineering 
\nref\KV{S. Katz, A. Klemm and C. Vafa,
``Geometric Engineering of Quantum Field Theories'',
\np{497}{1997}{173}, hep-th/9609239.}%
\nref\KVa{S. Katz and C. Vafa, ``Geometric Engineering of N=1
Quantum Field Theories'', \np{497}{1997}{196},
hep-th/9611090.}%
\nref\BJV{M. Bershadsky, A. Johansen, T. Pantev, V. Sadov and C. Vafa,
``F-theory, Geometric Engineering and N=1 Dualities'',
\np{505}{1997}{153}, hep-th/9612052.}%
\nref\KMV{S. Katz, P. Mayr and C. Vafa,
``Mirror symmetry and Exact Solution of 4D N=2 Gauge Theories I'',
hep-th/9706110.}%
\refs{\KV, \KVa, \BJV, \KMV}
following closely \IMS. 
Non-Abelian gauge theories in five dimensions arise from
compactifications of M-theory on singular Calabi-Yau manifolds. 
The singularity structure is studied in terms of the resolved
space $\pi:X\rightarrow{\bar{X}}$ which contains 
a collection of rationally ruled surfaces $S_j$ shrinking to a curve
${\bar C}$ in ${\bar X}$. Under the map $\pi$ the holomorphic curve
class $[\epsilon_j]$ of the ruling on  each surface shrinks to a point
in ${\bar C}$. Membranes wrapping the generic fibers yield BPS states
filling out a non-Abelian vector multiplet. More precisely, the 
simple roots of $G$ are identified with the divisor classes
$[S_i]\in H^2(X,{\bf Z})$ and the generic fibers
$[\epsilon_j]\in H_2(X,{\bf Z})$ with the simple co-roots. Their
intersection form reproduces the Cartan matrix of $G$\foot{We do not
differentiate in the text between the Cartan matrix its minus.
The correct signs appear in the formulas.}:
\eqn\cartan{S_i\cdot\epsilon_j=-C_{ij}.}
For simply laced groups, all divisors are ruled over curves of 
the same genus $g$ which determines the number of hypermultiplets
\ref\KMP{S. Katz,
D.R. Morrison and M.R. Plesser, ``Enhanced Gauge
Symmetry in Type II String Theory'', \np{477}{1996}{105},
hep-th/9601108.}.
In the non-simply laced case, the divisors corresponding to 
short simple roots intersect those corresponding to long roots 
along double or triple sections of genus $g^\prime$. The genus $g$ 
of a simple section determines the number of adjoint hypermultiplets.
As explained below, we also obtain $g^\prime -g$ extra hypermultiplets 
in a different representation. 

There are two distinct deformations corresponding
to charged matter. The first, generically associated to simply 
laced groups\foot{Note that there is an exception to this rule, namely
$Sp(N)$ gauge groups and fundamental representation.}, 
consists of blowing-up a given number of points on 
the minimal rulings. The resulting exceptional curves $\sigma_k$
account for weight vectors of certain group representations. 
More precisely, $H_2(X,{\bf Z})$ is identified with
the weight lattice of $G$. The intersection numbers of the 
holomorphic curves $\sigma_k$ with the exceptional divisors $S_i$
reproduce the weight system of the matter representation. 
All the other weight vectors are obtained by adding linear combinations
of the generic fiber classes to $\sigma_k$. The number of flavors is 
controlled by the number of independent exceptional curves,
or in some cases, the number of configurations of exceptional curves. 

The second deformation is characteristic to 
non-simply laced groups when one of the simple co-roots is already a
weight of a matter representation other than the adjoint.
The low energy theory then contains $g^\prime -g$ charged 
hypermultiplets. Note that in this case, a non-trivial matter 
content can be engineered with minimally ruled surfaces.
The ${\bf 7}$ representation of $G_2$ and the ${\bf 26}$
representation of $F_4$ are included in this category. 

Since the area of the curve ${\bar C}$ is inversely proportional to
the classical gauge coupling, a necessary and sufficient condition for
the existence of a fixed point is the existence of a contraction,
mapping ${\bar C}$ to a point. As shown in \IMS, this condition
implies that the prepotential is convex. A sufficient condition for
the existence of this contraction map is that the restriction to each
surface $S_j$ of a generic class in the relative \ka\ cone 
${\cal K}(X/{\bar X})$ is ample. Thus, one has to check that these 
restrictions have
positive intersection numbers with all irreducible holomorphic curves
on each $S_j$. We show that this condition is satisfied if the number
of flavors is low enough, thus establishing the existence of fixed point
theories. 


\newsec{$G_2,\, F_4$ and $E_8$}

As stated in the introduction, we begin the analysis with the simple 
cases which can be realized in terms of minimally ruled surfaces.
This is a known feature of non-simply laced groups noted in 
\refs{\IMS}. $E_8$ is included in the same category since the 
smallest matter representation is also the adjoint.

\subsec{$G_2$ with $n_7$ quarks}

Since $G_2$ has rank two, the corresponding degeneration consists of 
two rationally ruled surfaces $S_1,S_2$ over curves
$\gamma_1,\gamma_2$ where $\gamma_1$ is a triple cover 
of $\gamma_2$. In order to avoid generation of adjoint matter,
$\gamma_1$ is taken of genus zero while the genus of $\gamma_2$ 
determines the number of fundamental quarks $g=n_7$.

\ifig\gtwo{$G_2$ degeneration}{\epsfxsize2.0in\epsfbox{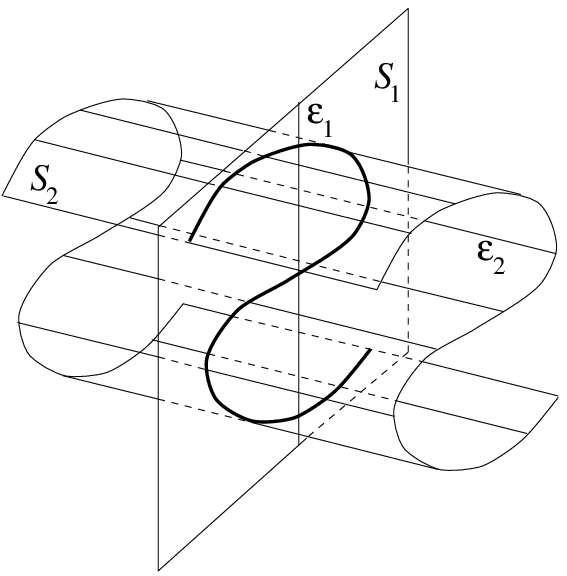}}

\noindent
Let $\epsilon_{1,2}$ denote the fiber classes of the rulings. 
The intersection matrix of the degeneration is the
$G_2$ Cartan matrix
\eqn\intersA{
\epsilon_i\cdot S_j=
\left[\matrix{-2 & 3 \cr
               1& -2\cr}\right].}
The matter content can be easily derived noting that the fiber class
$\epsilon_2$ is actually a weight vector for the ${\bf 7}$
representation. The other weight vectors can be realized as linear 
combinations of $\epsilon_{1,2}$. An arbitrary divisor supported 
on the exceptional locus can be written as 
\eqn\exdivA{
S=\phi_1 S_1+\phi_2 S_2}
where $\phi_{1,2}$ are coordinates on the negative relative \ka\ 
cone of the degeneration $-{\cal K}(X/{\bar X})$. The latter is defined
by 
\eqn\coneA{
-S\cdot \epsilon_i>0,\qquad i=1,2,}
therefore it corresponds to the fundamental Weyl chamber of $G_2$.
The prepotential is given by the triple intersection 
\eqn\prepA{
{\cal F}={1\over 6}S^3.}
Since the two surfaces intersect along a curve of genus $g$ in the
Calabi-Yau space, we have 
\eqn\trintA{
S_1^2S_2+S_1S_2^2=2g-2.}
As $\gamma$ is a 3-section in $S_1$,
$\gamma=3C^{\infty}+a\epsilon_1$ and
the adjunction formula shows that
\eqn\trisect{
\left(\gamma^2\right)_{S_1}=3(g+2).}
Therefore
\eqn\trintB{
S_1S_2^2=3(g+2),\qquad S_1^2S_2=-g-8}
and $S_1,S_2$ are ruled surfaces of degrees $n_1={1\over 3}(g+2(1-a))$
$n_2=g+8$. 
The final formula for the prepotential is 
\eqn\prepB{
{\cal F}=8\phi_1^3+8(1-g)\phi_2^3+9(g+2)\phi_1\phi_2^2
-3(g+8)\phi_1^2\phi_2}
which agrees with the gauge theory computation. 
Note that the result does not depend on the multiplicity $a$ of the 
fiber in the 3-section $\gamma$. However, this will turn out to be important
for the existence of fixed points. In principle, $a$ can be taken zero
unless a non-zero value is required by the integrality of $n_1$ and
$g$. In the present cases, we consider fixed values of $a$ as follows
\eqn\aval{
a=\left\{\matrix{& 1,\hfill &\qquad g\equiv 0(\hbox{mod} 3)\cr
                 & 0,\hfill &\qquad g\equiv 1(\hbox{mod} 3)\cr
                 & 2,\hfill &\qquad g\equiv 2(\hbox{mod} 3)\cr}
\right.}
This implies that $n_1\geq 0$. 

Next, we address the issue of the existence of UV fixed points. In
order to apply the method of \IMS\ based on contraction criteria
we need to explicitly find the extremal rays of the \ka\ cone
\coneA.\ This can be done systematically noting that the inequalities
\coneA\ can be rewritten in the form 
\eqn\coneB{
a_i>0,\qquad i=1,2}
by a linear change of variables 
\eqn\varchA{
a_i=C_{ij}\phi_j.}
Therefore the divisors on the extremal rays are determined by the 
columns of the $G_2$ quadratic form\foot{Strictly speaking,
$(C^{-1})_{ij}$ is the $G_2$ quadratic form with the second column
multiplied by 3. In the inequalities we consider this difference is
not important.}
\eqn\extrdivA{
L_i=\sum_{j}\left(C^{-1}\right)_{ij}S_j.}
Concretely,
\eqn\extrdivB{
L_1=2S_1+S_2,\qquad L_2=3S_1+2S_2.}
Note that this procedure of determining the extremal rays is general
and will be applied for all cases studied in this paper. In certain
situations, the resulting cone is divided into sub-cones corresponding 
to different geometric phases. This complication does not arise for 
minimally ruled configurations.

Following the strategy adopted in \IMS,\ we now check that the 
divisors $L_i$ lying on the extremal rays contract certain $S_j$ 
and induce ample classes on the resulting birational models. 
We have 
\eqn\intersC{\eqalign{
& -L_1\cdot S_1=C^{\infty}_1+\left({a\over 3}+
{2\over 3}(4-g)\right)\epsilon_1\cr
& -L_1\cdot S_2=(g+10)\epsilon_2\cr
& -L_2\cdot S_1=(4-g)\epsilon_1\cr
& -L_2\cdot S_2=C^{\infty}_2+(g+12)\epsilon_2.\cr}}
It follows that $L_1$ contracts $S_2$ along the ruling and it
restricts to an ample divisor on $S_1$ if $g<4$. If $g=4$, $L_1$
intersects $S_1\simeq {\f}_2$ along $C_1^{\infty}$. Therefore it 
also contracts $C^0_1$ yielding the cone ${\s}_1$ on which
$C^\infty_1$ is ample. 
Similarly,
$L_2$ contracts $S_1$ along the ruling if $g< 4$ and it is 
ample on $\s_2$. If $g=4$, $L_2$ contracts $S_1$ and the zero section 
on $S_2$ yielding the cone ${\s}_2$. Therefore, the class 
$-L_2\cdot S_2$ is ample
on $\s_2$.
We conclude that the theories exhibit UV fixed points
for $g\leq 4$. This is precisely the necessary condition derived in 
\IMS\ based on Higgsing arguments. The above analysis proves that it 
is is also sufficient. 

\subsec{$F_4$ with $n_{26}$ quarks}

The degeneration consists of a chain of four rationally ruled 
surfaces $S_1,\ldots, S_4$ intersecting along common curves
$\gamma_1,\ldots, \gamma_3$ as in fig. 2. $S_1$ and $S_2$ are 
ruled over rational curves while $S_3$ and $S_4$ are ruled over curves
of genus $g=n_{27}$. The curve $\gamma_2=S_2\cap S_3$ covers the rational 
base of $S_2$ twice. 

\ifig\Eeight{$F_4$ degeneration}{\epsfxsize2.5in\epsfbox{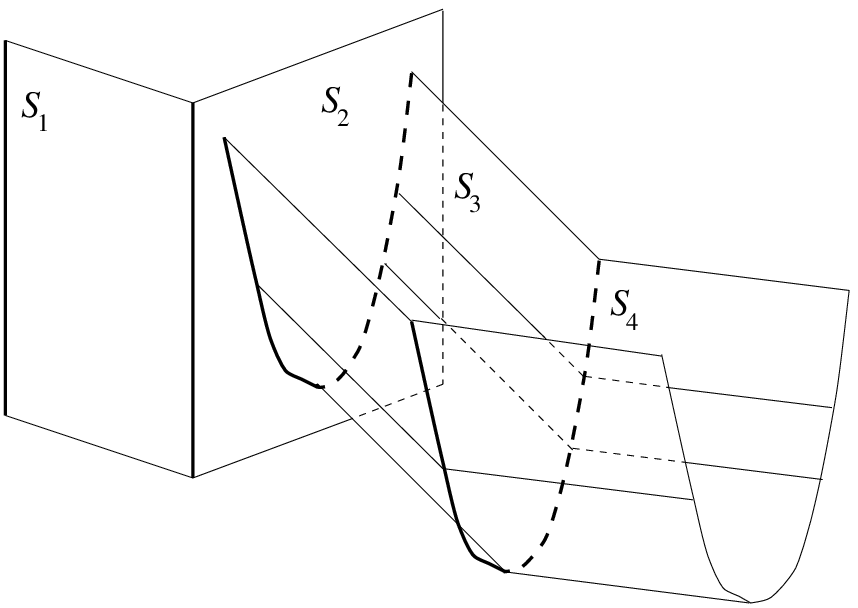}}

The intersection matrix 
\eqn\intersA{
\epsilon_i\cdot S_j=
\left[\matrix{-2 & 1 & 0 & 0\cr
               1 & -2 & 2 & 0\cr
               0 & 1 & -2 & 1\cr
               0 & 0 & 1 & -2\cr}
\right].}
reproduces correctly the $F_4$ Cartan matrix. 
The matter content can be easily determined by noting that all the weight
vectors of ${\bf 26}$ can be written as linear combinations of the
fiber classes $\epsilon_1,\ldots, \epsilon_4$. The relative \ka\ cone 
of the degeneration is defined by 
\eqn\coneC{
-S\cdot \epsilon_i>0,\qquad i=1,\ldots, 4}
where $S=\sum_{i=1}^4 \phi_i S_i$ is an arbitrary divisor supported
on the exceptional locus. These conditions define again the
fundamental Weyl chamber as expected from gauge theory considerations.
In order to compute the prepotential \prepA,\ we have to evaluate all
triple intersections of the form $S_i\cdot S_j\cdot S_k$.
Since $\gamma_2$ is a 2-section of $S_2$,
the adjunction formula yields
\eqn\adjB{
\left(\gamma_2\right)^2=4(g+1),\qquad n_2=g+1.}
Using \trintA,\ it follows that 
\eqn\trintC{\eqalign{
& S_1^2S_2=-g-1,\hskip 35pt S_1S_2^2=g-1\cr
& S_2^2S_3=-2g-6,\hskip 30pt S_2S_3^2=4(g+1)\cr
& S_3^2S_4=-8,\hskip 52pt S_3S_4^2=2g+6.}}
Note that these relations determine the degrees of the ruled surfaces
$S_1,\ldots,S_4$
\eqn\degffour{\eqalign{&n_1=g-1\hskip 35pt n_3=2g+6\cr
&n_2=g+1\hskip 35pt n_4=8.}}
Therefore the prepotential is given by 
\eqn\prepC{\eqalign{
{\cal F}=& 8\phi_1^3-3(g+1)\phi_1^2\phi_2+3(g-1)\phi_1\phi_2^2+8\phi_2^3
+12(g+1)\phi_2\phi_3^2
-3(2g+6)\phi_2^2\phi_3\cr 
& +8(1-g)\phi_3^3
+3(2g+6)\phi_3\phi_4^2-24\phi_3^2\phi_4+8(1-g)\phi_4^3.\cr}}
It can be checked by direct computation that this agrees with the 
one loop prepotential in gauge theory. 

According to the general procedure outlined above, the fixed point 
conditions can be expressed in terms of divisors $L_i,\ i=1,\ldots, 4$
lying on the extremal
rays of the relative \ka\ cone. These can be read off directly from
the $F_4$ quadratic form as in \extrdivA.
Next, we check that $L_i$ either contract certain $S_j$ or induce 
ample classes on the resulting birational models. 
Consider for example,
\eqn\extrdivC{
L_1=2S_1+3S_2+2S_3+S_4.}
We have 
\eqn\dintB{\eqalign{
& -L_1\cdot S_1=C_1^{\infty}+2(3-g)\epsilon_1\cr
& -L_1\cdot S_2=(5-g)\epsilon_2\cr
& -L_1\cdot S_3=2(5-g)\epsilon_3\cr
& -L_1\cdot S_4=2(5-g)\epsilon_4.\cr}}
This shows that $L_1$ contracts $S_2,S_3$ and $S_4$ along the ruling and 
reduces to an ample divisor on $S_1$ if $g<3$. If $g=3$, $L_1$ also 
contracts the zero section on $S_1$ and the induced class is ample 
on the resulting cone $\s_1$. 
The divisors corresponding to the remaining rays can be treated 
similarly. We conclude that a sufficient condition
for the existence of fixed points with $F_4$ gauge symmetry is
$n_{26}\leq 3$. Note that this is again precisely the necessary 
condition derived from a Higgsing argument in \IMS. 

\subsec{$E_8$ with $n_{248}$ quarks}

Since the matter multiplets transform in the adjoint representation,
the degeneration consists of a chain of rationally ruled surfaces
intersecting along sections according the $E_8$ Dynkin diagram. 
There are no multiple covers since $E_8$ is simply laced.
The number of adjoint quarks is equal to the genus of the common
base. Such a configuration results naturally from the resolution
of a curve of $E_8$ singularities in the Calabi-Yau manifold. 

\ifig\Eeight{$E_8$ degeneration}{\epsfxsize3.0in\epsfbox{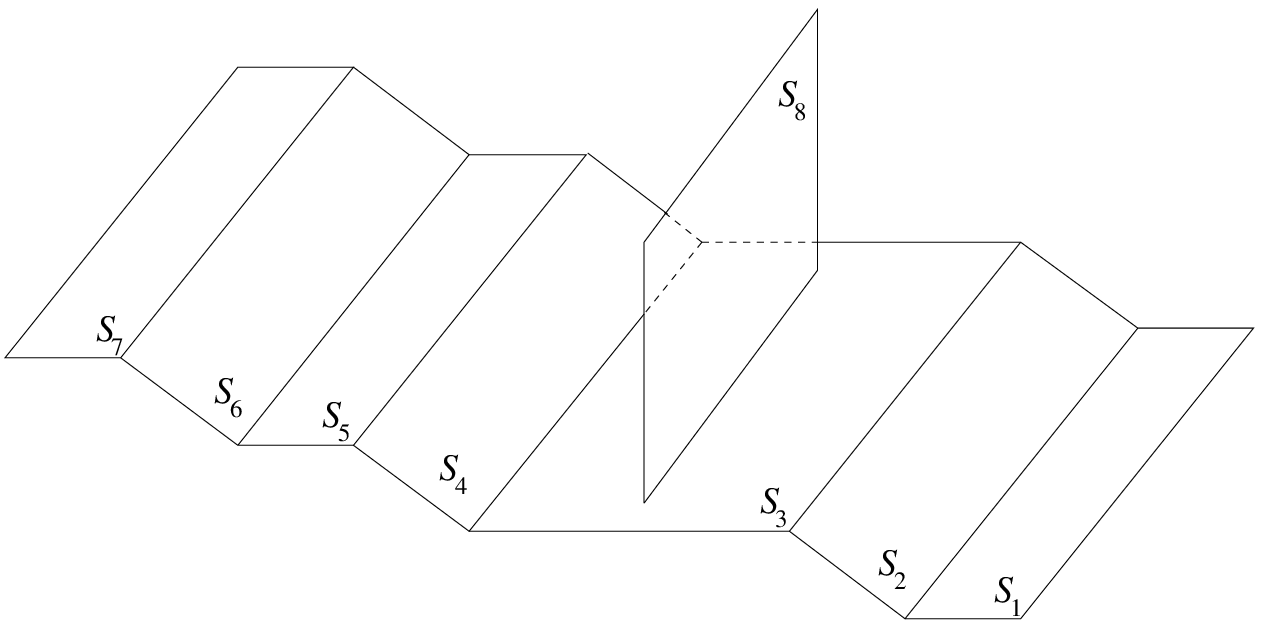}}

\noindent
It is straightforward to check that the intersection matrix 
$\epsilon_i\cdot S_j$ reproduces the $E_8$ Cartan matrix. 
The superpotential is given again by the triple intersection 
\prepA.\ The intersection numbers $S_i\cdot S_j\cdot S_k$ can be
computed starting from $S_3$. This minimally ruled surface 
must have three disjoint sections corresponding to the intersections
with $S_2,S_4,S_8$. The only way this can be realized is if $S_3$ is 
a surface of degree zero, therefore
\eqn\trintE{
S_3S_2^2=S_3S_4^2=S_3S_8^2=0.}
The remaining intersection numbers can be computed recursively
\eqn\trintF{\eqalign{
& S_2S_3^2=S_4S_3^2=S_8S_3^2=2g-2\cr
& S_1^2S_2=S_4S_5^2=2-2g\cr
& S_1S_2^2=S_4^2S_5=4g-4\cr
& S_5S_6^2=4-4g,\qquad S_5^2S_6=6g-6\cr
& S_6S_7^2=6-6g,\qquad S_6^2S_7=8g-8.\cr}}
The degrees of the rulings turn out to be 
\eqn\degeeight{\eqalign{
& n_1=n_5=4g-4\hskip 45pt n_6=6g-6\cr
& n_2=n_4=n_8=2g-2\hskip 23pt n_7=8g-8.\cr
& n_3=0}}
The prepotential resulting from \trintF\ is given by
\eqn\prepeight{\eqalign{{\cal F}&=8\phi_1^3+
18\phi_1^2\phi_2-24\phi_1\phi_2^2+8\phi_2^3+
12\phi_2^2\phi_3-18\phi_2\phi_3^2+8\phi_3^3+
6\phi_3^2\phi_4-12\phi_3\phi_4^2\cr&+8\phi_4^3-
6\phi_4\phi_5^2+8\phi_5^3-6\phi_5^2\phi_6+
8\phi_6^3-12\phi_6^2\phi_7+6\phi_6\phi_7^2+
8\phi_7^3-6\phi_5^2\phi_8+8\phi_8^3}}
and agrees with the gauge theory computation.

The fixed point analysis follows the same steps as in the previous
examples. Since there are no subtle points, we do not present the
details here. The result is that the divisors corresponding to
extremal rays contract most of the surfaces along a ruling or reduce
to ample classes on birational models if $g\leq 1$. The case 
$g=1$ is special as it corresponds to an $N=4$ gauge
theory. Geometrically, the Calabi-Yau threefold reduces to a 
direct product $T^2\times K3$. It is known that these theories 
cannot exhibit superconformal fixed points 
\ref\nahm{W. Nahm, ``Supersymmeries and Their Representations'', 
\np{135}{1978}{149}.}.
Therefore, there is a unique fixed point with $E_8$ gauge symmetry 
corresponding to the theory without matter.

\newsec{$E_6$ with $n_{27}$ quarks}

\subsec{Degenerations and phase structure}

A systematic procedure for constructing $E_6$ degenerations with
matter is to start from an F-theory compactification on a singular
Weierstrass model. The type of singular elliptic fibers and the 
associated matter content have been classified in 
\nref\MV{D.R. Morrison and C. Vafa, ``Compactifications of F-Theory on 
Calabi--Yau Threefolds -- I'', \np{473}{1996}{74}, 
hep-th/9602114; ``Compactifications of F-Theory on Calabi--Yau 
Threefolds -- II'', \np{476}{1996}{437}
hep-th/9603161.}
\nref\BIK{M. Bershadsky, K. Intriligator, S. Kachru, D.R. Morrison,
V. Sadov and C. Vafa, ``Geometric Singularities and Enhanced Gauge
Symmetries'', \np{481}{1996}{215}, hep-th/96050200.}%
\refs{\MV,\BIK}.
The strategy is to construct a smooth elliptic model as in 
\nref\RM{R. Miranda, ``Smooth Models for Elliptic Threefolds'', in
R. Friedman and D.R. Morrison, editors, ``The Birational Geometry of
Degenerations'', Birkhauser, 1983.}%
\nref\AG{P.S. Aspinwall and M. Gross, ``The $SO(32)$ Heterotic String
on a K3 Surface'',  Phys. Lett. {\bf B387} (1996) 735, hep-th/9605131.}%
\nref\A{P.S. Aspinwall, ``Point-Like Instantons and the
$Spin(32)/Z_2$ Heterotic String'', \np{496}{1997}{149},
hep-th/9612108.}%
\nref\AM{P.S. Aspinwall, D.R. Morrison, ``Point-like Instantons on 
K3 Orbifolds'', \np{503}{1997}{533}, hep-th/9705104.}
\refs{\RM,\AG,\A, \AM}.
Finally, we can take the M-theory limit by sending the size of
the original elliptic fiber to infinity. We explicitly carry out 
this procedure for $E_6$ in appendix B. The result is presented
in the figure below. 

\ifig\Esixone{$E_6$ degeneration - phase I}
{\epsfxsize4.0in\epsfbox{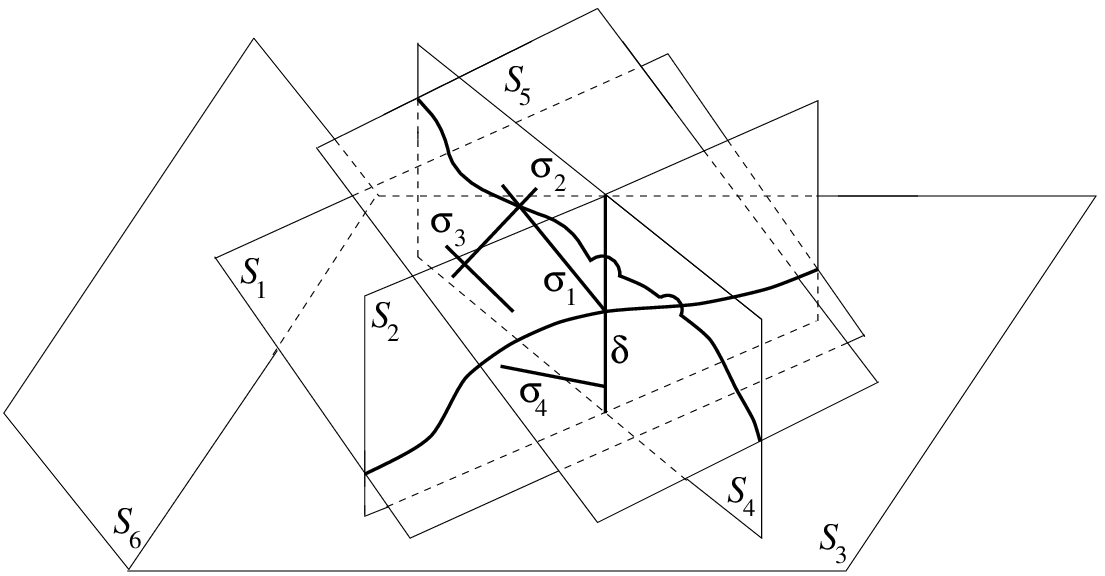}}

\noindent
The surfaces $S_2, S_3, S_5$ and $S_6$ are minimally ruled, while the
surfaces $S_1, S_4$ are blown-up twice so that they each contain 
a reducible fiber with three components
\eqn\fibersixone{\eqalign{
\epsilon_1&=\sigma_1+\sigma_2+\sigma_3,\cr
\epsilon_4&=\sigma_1+\delta+\sigma_4.}}
All surfaces are ruled over rational curves. 
Note that $S_2$ and $S_4$ intersect along the curve $\delta$ 
which is a fiber of $S_2$ and a $(-2)$ curve on $S_4$. Similarly,
$S_1$ and $S_5$ intersect along $\sigma_2$ which is a fiber in the
ruling of $S_5$ and a $(-2)$ curve on $S_1$. $S_1$ and $S_4$ intersect
along $\sigma_1$ which is an exceptional $(-1)$ curve in both
surfaces. Next, 
$S_5$ intersects $S_4$ 
along a section passing through the $(-1)$ curve $\sigma_1$
and $S_2$ intersects $S_1$ along a similar section.  
The intersection matrix of $\sigma_1,\ldots, \sigma_4$ and $\delta$ with 
the exceptional divisors is given by
\eqn\intmatsixone{\vbox{
\offinterlineskip
\halign{\strut
\quad $#$\quad &\vrule\hfill\quad $#$\quad\hfill&
\vrule\quad $#$\quad\hfill&\vrule\hfill\quad $#$
\quad\hfill&\vrule\hfill\quad $#$\quad\hfill&
\vrule\hfill\quad $#$\quad\hfill\cr
&S_1&S_2&S_3&S_4&S_5\cr
\noalign{\hrule}
\sigma_1&-1&1&0&-1&1\cr
\noalign{\hrule}
\sigma_2&0&0&0&1&-2\cr
\noalign{\hrule}
\sigma_3&-1&0&0&0&1\cr
\noalign{\hrule}
\sigma_4&0&1&0&-1&0\cr
\noalign{\hrule}
\delta&1&-2&1&0&0\cr
}}}
It can be easily checked that \fibersixone\ and \intmatsixone\ reproduce
the $E_6$ Cartan matrix. Note also that $\sigma_1,-\sigma_3$ and $-\sigma_4$
are weight vectors of the ${\bf 27}$ representation. The other weights
can be written as linear combinations of $\sigma_{1,3,4}$ and the 
fiber classes. The relative \ka\ cone corresponding to the present 
degeneration is determined by 
\eqn\relcone{\eqalign{
-S\cdot\epsilon_i&>0,\cr
-S\cdot\sigma_k&>0,\qquad -S\cdot\delta>0}}
where $D=\sum_{i=1}^6\phi_iS_i$.
To make the fixed point analysis later on easier, it is more
convenient to work in the coordinates
\eqn\coord{
a_i=C_{ij}\phi_j,}
where the external cone determined by the first set of inequalities is the
positive 'quadrant' $a_i>0$.
The $\sigma_{3,4}$ inequalities define a sub-cone\foot{Note that the
$\sigma_1$ inequality is redundant.} which is the intersection
of the cones 
\eqn\sixone{\eqalign{
\phi_5&<\phi_1,\cr
\phi_2&<\phi_4}}
with the first quadrant. We call the phase corresponding to this
sub-cone phase I. The form of the inequalities \sixone\ in the $a_i$
coordinates can easily be read off by taking the corresponding rows of
the $E_6$ quadratic form
\eqn\subconesix{\eqalign{2a_1+a_2-a_4-2a_5&>0,\cr
2a_4+a_5-a_1-2a_2&>0.}}

The guiding principle for uncovering the rest of the phases is finding
the degenerations which invert the inequalities \sixone. Flopping the
exceptional curves $\sigma_k$ from one surface to another changes their
intersection matrix with the divisors $S_i$. This leads to new
inequalities which further subdivide the relative \ka\ cone
and hence give rise to additional phases. As a consistency check one
has to verify that all phases are mutually disjoint and that their
union is exactly the extended \ka\ cone determined by the first
inequalities in \relcone. 

On the boundaries of the different sub-wedges charged hypermultiplets are
going to zero mass. This means that even though the order by which we 
proceed to invert the inequalities \sixone\ is not unique, the phase
structure that we find is. Indeed, in some cases it is very difficult
to see in the geometry the flops relating one degeneration to another
one. However, by checking the consistency of this scheme as outlined
above, we can be reasonably sure that the phase structure we describe 
is the correct one.

${\underline{\hbox{Phase II}}}$ -
The diagram describing this degeneration is identical to \Esixone,
but with $\sigma_3$ flopped from $S_1$ to $S_5$. It follows
that the reducible fibers are now given by
\eqn\fibersixtwo{\eqalign{
\epsilon_1&=\sigma_1+\sigma_2\cr
\epsilon_4&=\sigma_1+\delta+\sigma_4\cr
\epsilon_5&=\sigma_2+\sigma_3.}}
The intersection matrix with the exceptional divisors is given by
\eqn\intmatsixtwo{\vbox{
\offinterlineskip
\halign{\strut
\quad $#$\quad &\vrule\hfill\quad $#$\quad\hfill&
\vrule\quad $#$\quad\hfill&\vrule\hfill\quad $#$
\quad\hfill&\vrule\hfill\quad $#$\quad\hfill&
\vrule\hfill\quad $#$\quad\hfill\cr
&S_1&S_2&S_3&S_4&S_5\cr
\noalign{\hrule}
\sigma_1&-1&1&0&-1&1\cr
\noalign{\hrule}
\sigma_2&1&0&0&1&-1\cr
\noalign{\hrule}
\sigma_3&1&0&0&0&-1\cr
\noalign{\hrule}
\sigma_4&0&1&0&-1&0\cr
\noalign{\hrule}
\delta&1&-2&1&0&0\cr
}}}
As before, and in all other phases below, this intersection matrix
correctly reproduces the  Cartan matrix and the curves which are not
simple co-roots are weight vectors of the ${\bf 27}$ representation.
The inequalities which define this phase can be read off from \intmatsixtwo:
\eqn\sixtwo{\eqalign{
\phi_1&<\phi_5\cr
\phi_2&<\phi_4\cr
\phi_4&<\phi_1+\phi_5\cr
\phi_2+\phi_5&<\phi_1+\phi_4
.}}
Note that the $\sigma_1$ inequality is no longer redundant and will
therefore have to be inverted.

${\underline{\hbox{Phase III}}}$ -
The degeneration corresponding to this phase is described in fig. 5 below.
\ifig\Esixthree{$E_6$ degeneration - phase III}{\epsfxsize4.0in
\epsfbox{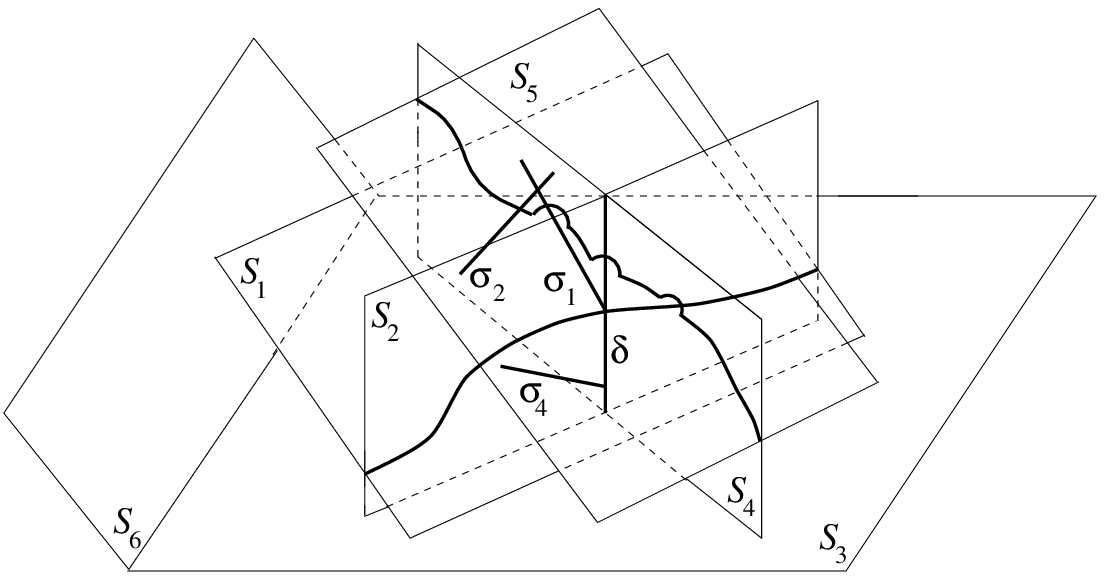}}

\noindent
All the surfaces are ruled over rational curves and for $S_4$ the
ruling is minimal. The $S_4$ fiber has four components and is given
\eqn\fiberthree{\epsilon_4=\sigma_1+\delta_4+\sigma_4+\sigma_2.}
$S_1$ and $S_4$ intersect along $\sigma_1$ which is the fiber in the
ruling of $S_1$. $S_2$ and $S_4$ intersect along $\delta$ which is
the fiber in the ruling of $S_2$. The section of $S_4$ along which it
intersects $S_5$ passes through $\sigma_2$ and avoids $\delta$ and
$\sigma_1$. The intersection matrix of these curves with the surfaces
$S_j$ is therefore given by
\eqn\intmatsixthree{\vbox{
\offinterlineskip
\halign{\strut
\quad $#$\quad &\vrule\hfill\quad $#$\quad\hfill&
\vrule\quad $#$\quad\hfill&\vrule\hfill\quad $#$
\quad\hfill&\vrule\hfill\quad $#$\quad\hfill&
\vrule\hfill\quad $#$\quad\hfill\cr
&S_1&S_2&S_3&S_4&S_5\cr
\noalign{\hrule}
\sigma_1&-2&1&0&0&0\cr
\noalign{\hrule}
\sigma_2&1&0&0&-1&1\cr
\noalign{\hrule}
\sigma_4&0&1&0&-1&0\cr
\noalign{\hrule}
\delta&1&-2&1&0&0\cr
}},}
leading to the following inequalities defining phase III
\eqn\sixthree{\eqalign{
\phi_2&<\phi_4\cr
\phi_1+\phi_5&<\phi_4.}}
The inequalities above imply
\eqn\ineqthree{\eqalign{\phi_1&<\phi_5\cr
\phi_2+\phi_5&<\phi_1+\phi_4,}}
where the first one follows from the second inequality in \sixthree\
by using the $\epsilon_5$ inequality in \relcone\ and the second
follows again from the second inequality in \sixthree\ by adding
$\phi_1$ to both sides and using the $\epsilon_1$ inequality in
\relcone. This shows that phase III is disjoint from phase I.

${\underline{\hbox{Phase IV}}}$ -
The degeneration corresponding to this phase is identical to that
described in \Esixone\ except that $\sigma_2$ is flopped from $S_4$ to
$S_2$. The non-minimally ruled surfaces are $S_1,S_2$ and $S_4$ whose 
reducible fibers are given by
\eqn\fiberfour{\eqalign{
\epsilon_1&=\sigma_1+\sigma_2+\sigma_3\cr
\epsilon_2&=\delta+\sigma_4\cr
\epsilon_4&=\delta+\sigma_1.}}
The intersection matrix is given by
\eqn\intmatsixfour{\vbox{
\offinterlineskip
\halign{\strut
\quad $#$\quad &\vrule\hfill\quad $#$\quad\hfill&
\vrule\quad $#$\quad\hfill&\vrule\hfill\quad $#$
\quad\hfill&\vrule\hfill\quad $#$\quad\hfill&
\vrule\hfill\quad $#$\quad\hfill\cr
&S_1&S_2&S_3&S_4&S_5\cr
\noalign{\hrule}
\sigma_1&-1&1&0&-1&1\cr
\noalign{\hrule}
\sigma_2&0&0&0&1&-2\cr
\noalign{\hrule}
\sigma_3&-1&0&0&0&1\cr
\noalign{\hrule}
\sigma_4&0&-1&0&1&0\cr
\noalign{\hrule}
\delta&1&-1&1&-1&0\cr
}},}
leading to the inequalities
\eqn\sixfour{\eqalign{
\phi_5&<\phi_1\cr
\phi_4&<\phi_2\cr
\phi_2+\phi_5&<\phi_1+\phi_4\cr
\phi_1+\phi_3&<\phi_2+\phi_4.}}
It is clear that this phase is disjoint from all the previous ones.

${\underline{\hbox{Phase V}}}$ -
The present degeneration is described in fig. 6 below.
\ifig\Esixfive{$E_6$ degeneration - phase V}{\epsfxsize4.0in
\epsfbox{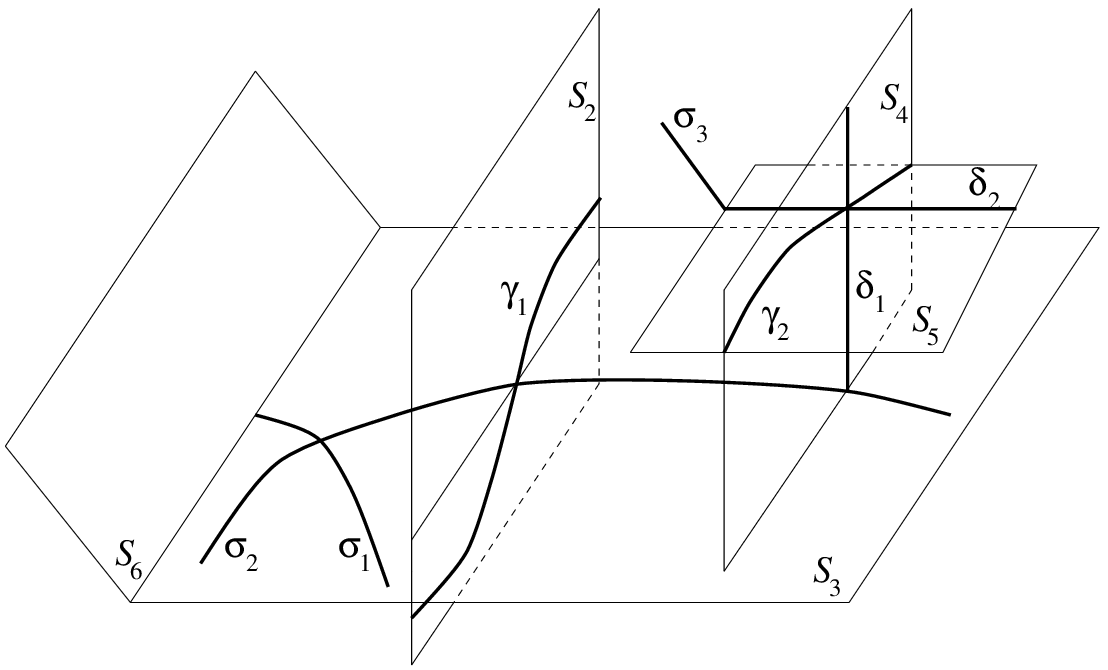}}

\noindent
The non-minimally ruled surfaces are $S_3$ and $S_1$ which intersects
$S_2$ along $\gamma_1$. This surface is too complicated to draw, but we
encourage the reader to use her or his imagination. The irreducible
fibers 
are given by
\eqn\fiberfive{\eqalign{
\epsilon_1&=\sigma_2+\delta_1+\delta_2+\sigma_3\cr
\epsilon_3&=\sigma_1+\sigma_2.}}
The intersection matrix is given by
\eqn\intmatsixfive{\vbox{
\offinterlineskip
\halign{\strut
\quad $#$\quad &\vrule\hfill\quad $#$\quad\hfill
&\vrule\hfill\quad $#$\quad\hfill&
\vrule\quad $#$\quad\hfill&\vrule\hfill\quad $#$
\quad\hfill&\vrule\hfill\quad $#$\quad\hfill&
\vrule\hfill\quad $#$\quad\hfill\cr
&S_1&S_2&S_3&S_4&S_5&S_6\cr
\noalign{\hrule}
\sigma_1&1&0&-1&0&0&1\cr
\noalign{\hrule}
\sigma_2&-1&1&-1&1&0&0\cr
\noalign{\hrule}
\sigma_3&-1&0&0&0&1&0\cr
\noalign{\hrule}
\delta_1&0&0&1&-2&-1&0\cr
\noalign{\hrule}
\delta_2&0&0&0&0&1&-2\cr
}}}
The inequalities defining this phase are therefore
\eqn\sixfive{\eqalign{
\phi_2+\phi_4&<\phi_1+\phi_3\cr
\phi_5&<\phi_1\cr
\phi_1+\phi_6&<\phi_3.}}
It is straight forward to verify that these imply
\eqn\ineqfive{\eqalign{
\phi_4&<\phi_2\cr
\phi_2+\phi_5&<\phi_1+\phi_4,}}
which shows that this phase is disjoint from phases I, II and III. By 
the first inequality in \sixfive, it is manifestly disjoint from phase
IV as well.

${\underline{\hbox{Phases VI and VII}}}$ -
These are obtained by flopping $\sigma_1$, first from $S_3$ to
$S_6$ and then outside $S_6$ as described in the figure 7 below.
\ifig\Esixsix{$E_6$ degeneration - phases VI and VII}{\epsfxsize4.0in
\epsfbox{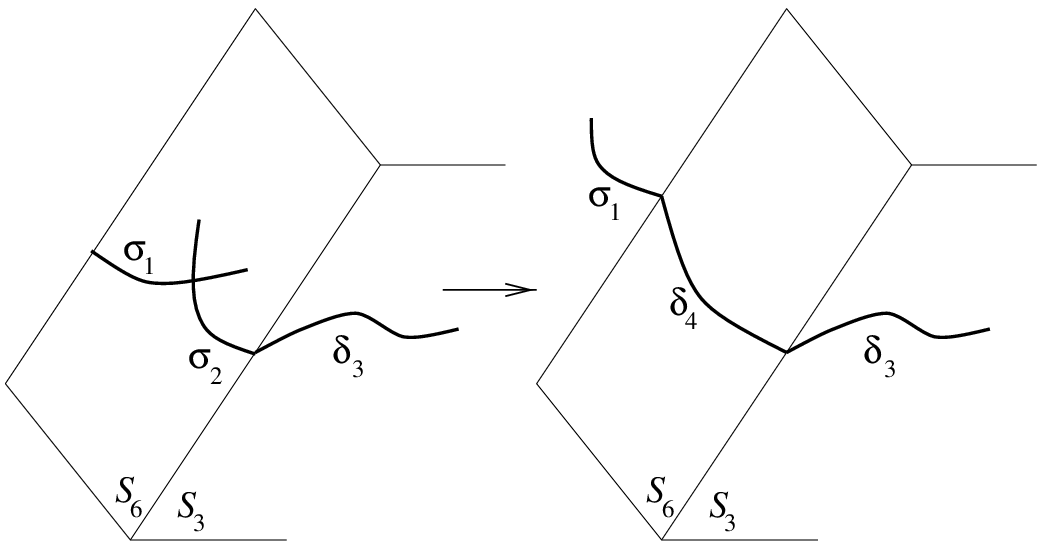}}

\noindent
The changes from the degeneration corresponding to phase V are
\eqn\diffsix{\vbox{
\offinterlineskip
\halign{\strut
\quad $#$\quad&\vrule\quad $#$\quad\hfill
&\vrule\quad $#$\quad\hfill\cr
&\hbox{phase VI}&\hbox{phase VII}\cr
\noalign{\hrule}
\epsilon_1&{\eqalign{&\sigma_3+\delta_1+\delta_2\cr&+\delta_3+\sigma_1}}
&{\eqalign{&\sigma_3+\delta_1+\delta_2\cr&+\delta_3+\delta_4+\sigma_1}}\cr
\noalign{\hrule}
\epsilon_3&\delta_3&\delta_3\cr
\noalign{\hrule}
\epsilon_6&\sigma_1+\sigma_2&\delta_4\cr
}}}
The different intersection of the fibers
above yield the inequalities
\eqn\sixsix{\eqalign{\hbox{phase VI}&\qquad\phi_3<\phi_1+\phi_6\cr
&\qquad\phi_1<\phi_6\cr
\hbox{phase VII}&\qquad\phi_6<\phi_1.}}
It can be easily checked that in both these phases the first two
inequalities defining phase V in \sixfive\ hold. Also, the inequality
defining phase VII implies $\phi_3<\phi_1+\phi_6$ showing that it is
disjoint from VI.

So far we have constructed seven phases. The reflection symmetry of
the $E_6$ Dynkin diagram means that
for every phase found so far there exists another {\it distinct} phase in
which $S_1$ is exchanged with $S_5$ and $S_2$ is exchanged with
$S_4$. We will denote these as $\phase{I},\phase{II},\ldots$. 
The defining inequalities of $\phase{I},\phase{II},\ldots$ are 
obtained by making a $2\leftrightarrow4$ and $1\leftrightarrow5$ 
exchange in the those defining the I,II,$\ldots$ phases. 
It is easily verified that the phases I, IV-VII, $\phase{II}$ and
$\phase{III}$ completely cover the $\phi_5<\phi_1$ region. It follows
by symmetry that the complement of this region is covered by
$\phase{I}$, II, III and $\phase{IV}$-$\phase{VII}$. We conclude
therefore that all possible inequalities have been exhausted and the
14 disjoint phases found completely cover the Coulomb phase.

\subsec{A prepotential calculation}

The prepotential can be computed in each phase by evaluating all
triple intersections of the form $S_i\cdot S_j\cdot S_k$. We carry 
this out in detail for the phase I defined in \sixone, all other cases
being similar.

In order to obtain $n_{27}$ quarks the divisors $S_2$ and $S_4$ must 
intersect $S_3$ along two sections, $\gamma_1$ and $\gamma_2$, meeting
transversely $n=n_{27}$ times. Since $S_3$ is minimally ruled, it must
be isomorphic to $\f_n$ and $\gamma_1\simeq\gamma_2\simeq
C^{\infty}_3$. It follows that $S_6$ intersects $S_3$ along
$C^0_3$. This results in the intersection numbers
\eqn\intsixone{\eqalign{
& S_3S_2^2=S_3S_4^2=n,\qquad\hskip 25pt S_3S_6^2=-n\cr
& S_3^2S_2=S_3^2S_4=-n-2,\qquad S_3^2S_6=n-2,}}
showing that 
\eqn\degsixone{
n_2=n_4=n+2,\qquad n_6=n-2.}
Note that $S_4$ is obtained from $\f_{n+2}$ by blowing-up $2n$ points 
lying at the intersection of $n$ fibers with two distinct $C^\infty_4$
sections. The strict transforms of the two sections are 
the holomorphic irreducible 
curves $C_4^\infty-\sum_{\alpha=1}^n\sigma^\alpha_1$ and
$C_4^\infty-\sum_{\alpha=1}^n\sigma^\alpha_3$
\ref\Ha{R. Hartshorne, ``Algebraic Geometry'', Graduate Texts in 
Mathematics, Springer-Verlag (1993).}.
The $S_5$ surface
intersects $S_4$ along the former, therefore
\eqn\intsixtwo{S_4S_5^2=2,\qquad S_4^2S_5=-4.\qquad n_5=4.}
The other non-minimal surface is $S_1$ which is blown-up at $2n$
points lying at the intersection of $n$ fibers with two sections
$C^0_1$ and $C^\infty_1$. Therefore it intersects $S_2$ along
the strict transform $C_1^0-=\sum_{\alpha=1}^n\sigma_1^\alpha$ 
in $S_1$ and along
$C_2^\infty$ in $S_2$. It follows that
\eqn\intnoB{
S_1^2S_2=n+2,\qquad S_1S_2^2=-n-4,\qquad n_1=4.}
Recall that 
\eqn\dbint{
S_2\cdot S_4=\sum_{\alpha=1}^n\delta^{\alpha},\qquad
S_1\cdot S_4=\sum_{\alpha=1}^n\sigma_1^{\alpha},\qquad
S_1\cdot S_5=\sum_{\alpha=1}^n\sigma_2^{\alpha},}
which determines
\eqn\intsixfour{\eqalign{
& S_2^2S_4=S_1S_5^2=-2n,\cr
&S_2S_4^2=S_1^2S_5=0,\cr
& S_1S_4S_5=S_1S_2S_4=S_2S_3S_4=n.}}
Finally, the remaining intersection numbers are 
\eqn\intsixfive{
S_2^3=S_3^3=S_5^3=S_6^3=8,\qquad S_1^3=S_4^3=8-2n.}
The prepotential, in this phase, determined by the intersection numbers
\intsixone-\intsixfive\ is 
given by
\eqn\prepesix{\eqalign{{\cal
F}&=(8-2n)\phi_1^3+3(n+2)\phi_1^2\phi_2-3(n+4)
\phi_1\phi_2^2+8\phi_2^3+3n\phi_2\phi_3^2-3(n+2)\phi_2\phi_3^2\cr
&+8\phi_3^3-3n\phi_1^2\phi_4+6n\phi_1\phi_2\phi_4-6n\phi_2^2
\phi_4+6n\phi_2\phi_3\phi_4-3(n+2)\phi_3^2\phi_4-3n\phi_1\phi_4^2\cr
&+3n\phi_3\phi_4^2+(8-2n)\phi_4^3+6n\phi_1\phi_4\phi_5-12\phi_4^2
\phi_5-6n\phi_1\phi_5^2+6\phi_4\phi_5^2+8\phi_5^3\cr
&+3(n-2)\phi_3^2\phi_6-3n\phi_3\phi_6^2+8\phi_6^3
}}
and agrees with the gauge theory result. 

\subsec{Fixed points analysis}

The fixed point analysis is more involved than the previous cases
and will be presented in detail. As shown in the first sub-section, 
the extended \ka\ cone is divided into sub-cones, each corresponding
to a smooth  model. In each phase, we can contract the fibers and
exceptional fiber components of the $S_i$ obtaining a singular
threefold ${\bar X}$ with a curve of singularities ${\bar C}$ as in \IMS.\
Note that ${\bar C}$ is the same for all phases. Since the
contraction criterion establishes whether the curve ${\bar C}$ can be further
contracted in ${\bar X}$, it is enough to check it in a single phase
which can be chosen arbitrarily. In the following we focus on the 
smooth model of phase I. 

Consider a generic divisor
\eqn\gendivA{
S=\sum_{i=1}^6\phi_iS_i.}
In phase I given by \sixone, the restrictions of $S$ to the surfaces
$S_i$ are given by
\eqn\dbintC{\eqalign{
H_1\equiv -S\cdot S_1=\ & (2\phi_1-\phi_2)C^{\infty}_1+
(4\phi_2-2\phi_1-n\phi_5)\epsilon_1\cr
&-(\phi_1+\phi_4-\phi_2-\phi_5)\sum_{\alpha=1}^n\sigma_1^{\alpha}
 -(\phi_1-\phi_5)\sum_{\alpha=1}^n\sigma_3^\alpha\cr
H_2\equiv -S\cdot S_2=\ & (2\phi_2-\phi_1-\phi_3)C^{\infty}_2+
(2\phi_3-n(\phi_2+\phi_4-\phi_3))\epsilon_2\cr
H_3\equiv -S\cdot S_3=\ & (2\phi_3-\phi_2-\phi_4-\phi_6)C^{\infty}_3+
(2\phi_3-n(\phi_3-\phi_6))\epsilon_3\cr
H_4\equiv -S\cdot S_4=\ & (2\phi_4-\phi_3-\phi_5)C^{\infty}_4+
(2\phi_3-n(\phi_2+\phi_4-\phi_3))\epsilon_4\cr
& -(\phi_1+\phi_4-\phi_2-\phi_5)\sum_{\alpha=1}^n\sigma_1^{\alpha}
-(\phi_4-\phi_2)\sum_{\alpha=1}^n\sigma_4^\alpha\cr
H_5\equiv -S\cdot S_5=\ & (2\phi_5-\phi_4)C^{\infty}_5+
(4\phi_4-2\phi_5-n\phi_1)\epsilon_5\cr
H_6\equiv-S\cdot S_6=\
&(2\phi_6-\phi_3)C^{\infty}_6+(4-n)\phi_6\epsilon_6.}}

As explained before, we have to check that for a generic divisor $S$ the
induced classes $H_i$ are ample, and therefore must have positive
intersection number with all irreducible holomorphic curves 
on $S_i$. We will actually check that $H_i$ are positive with respect
to all effective divisors on $S_i$. 
We start with the minimally ruled
surfaces $S_2$, $S_3$, $S_5$ and $S_6$. In these cases the $H_i$ are
ample if and only if the coefficients of $C^\infty_i$ and $\epsilon_i$
are positive \Ha.
Taking into account the
conditions \sixone,\ it follows that $n$ must satisfy the bound 
$n\leq 3$. Note that all surfaces except $S_6$ actually allow $n\leq
4$. 

Next we consider the blown-up surface $S_1$. Since the Picard group of
$S_1$ is generated by $C^0_1,
\epsilon_1,\sigma_1^\alpha$ and $\sigma_3^\alpha$, 
an arbitrary divisor class can be written as 
\eqn\geneffone{D_1=a_1C_1^0+b_1\epsilon_1+
\sum_{\alpha=1}^nc_\alpha\sigma_1^\alpha
+\sum_{\alpha=1}^nd_\alpha\sigma_3^\alpha.}
As showed in 
\nref\Dem{M. Demazure, in ``S\'eminaire sur les Singularit\'es des Surfaces'',
(M. Demazure, H. Pinkham and B. Teissier - eds.),
Lecture Notes in Mathematics {\bf 777} Springer-Verlag, 1980, p. 24.}%
\nref\Reid{M. Reid, ``Chapters on Algebraic Surfaces'',
alg-geom/9602006.}%
\refs{\Dem,\Reid},
if the class \geneffone\ has an effective representative, it must 
have non-negative intersection with all irreducible holomorphic curves
$C$ such that $C^2\geq 0$. In particular, we can consider $C$ to be
one of the following 
\eqn\curvesone{C_1^\infty,\quad \epsilon_1,\quad
C_1^\infty-\sum_{\alpha=1}^n\sigma_3^\alpha
,\quad C_1^\infty+n\epsilon_1-\sum_{\alpha=1}^n\sigma_3^\alpha.}
which yield the conditions
\eqn\condone{\eqalign{
&a_1\geq0\cr
&b_1\geq0\cr
&b_1+\sum_{\alpha=1}^nd_\alpha\geq0\cr
&b_1+na_1+\sum_{\alpha=1}^nc_\alpha\geq0.
}}
To check the ampleness of the class $H_1$ we rewrite $D$ as 
\eqn\geneftwo{\eqalign{
D&=a_1\left(C^0_1-\sum_{\alpha=1}^n\sigma_1^\alpha\right)+b_1
\left(\sigma_1+\sigma_2+\sigma_3\right)+
\sum_{\alpha=1}^n(a_1+c_\alpha)\sigma_1^\alpha+\sum_{\alpha=1}^n
d_\alpha\sigma_3^\alpha,}}
which gives
\eqn\ampone{\eqalign{H_1\cdot D_1
&=a_1\left(C_1^0-\sum_{\alpha=1}^n\sigma_1^\alpha\right)\cdot
H_1+\left(b_1+na_1+\sum_{\alpha=1}^nc_\alpha\right)H_1\cdot\sigma_1\cr
&+\left(b_1+\sum_{\alpha=1}^nd_\alpha\right)H_1\cdot\sigma_3+b_1H_1
\cdot\sigma_2.}}
It can be checked by direct computation that inside the sub-cone
\sixone, all the intersections appearing in \ampone\ are positive for
$n\leq4$. Therefore, using the conditions \condone, we find that $H_1$
restricts to an ample class on $S_1$ when $n\leq4$. 

Finally, we consider the class $H_4$. A generic divisor on
$S_4$ is given by 
\eqn\geneffthree{D_4=a_4C^0_4+b_4\epsilon_4+\sum_{\alpha=1}^nc_\alpha\sigma_1^\alpha+\sum_{\alpha=1}^nd_\alpha\sigma_4^\alpha.}
The conditions on the coefficients imposed by requiring the class
\geneffthree\ to have an effective representative are
\eqn\condtwo{\eqalign{
&a_4\geq0\cr
&b_4\geq0\cr
&b_1+\sum_{\alpha=1}^nc_\alpha\geq0\cr
&b_1+\sum_{\alpha=1}^nd_\alpha\geq0.
}}
These are derived using the holomorphic, positive self-intersection
curves
\eqn\curvestwo{C_4^\infty,\quad\epsilon_4,\quad
C_4^\infty-\sum_{\alpha=1}^n\sigma^\alpha_1,\quad 
C_4^\infty-\sum_{\alpha=1}^n\sigma^\alpha_4.}
After a similar manipulation as above, we get
\eqn\amptwo{H_4\cdot D_4=a_4 C_0\cdot
H_4+\left(b_4+\sum_{\alpha=1}^nc_\alpha\right)\sigma_1\cdot
H_4+\left(b_4+\sum_{\alpha=1}^nd_\alpha\right)\sigma_4\cdot
H_4+b_4\delta\cdot H_4.}
As before, it can be easily shown that for classes $H_4$ inside the sub-cone
\sixone\ all the intersections in \amptwo\ are positive if $n\leq4$. 
Together with the condition \condtwo\ this implies that $H_4$ restricts
to an ample class on $S_4$ if $n\leq 4$.

The end result is that the classes $H_i=-S\cdot S_i$ are ample for 
any divisor $S$ inside the sub-cone if $n\leq 3$. We conclude that 
for these values of $n=n_{27}$ there exist UV fixed points with 
$E_6$ gauge symmetry.

\newsec{$E_7$ with $n_{56}$ quarks}

\subsec{Degenerations and phase structure}
An $E_7$ degeneration can be similarly constructed by performing 
canonical resolution of the corresponding F-theory Weierstrass model
(appendix C). The result is represented below. 

\ifig\Esevenone{$E_7$ degeneration - phase I}
{\epsfxsize4.0in\epsfbox{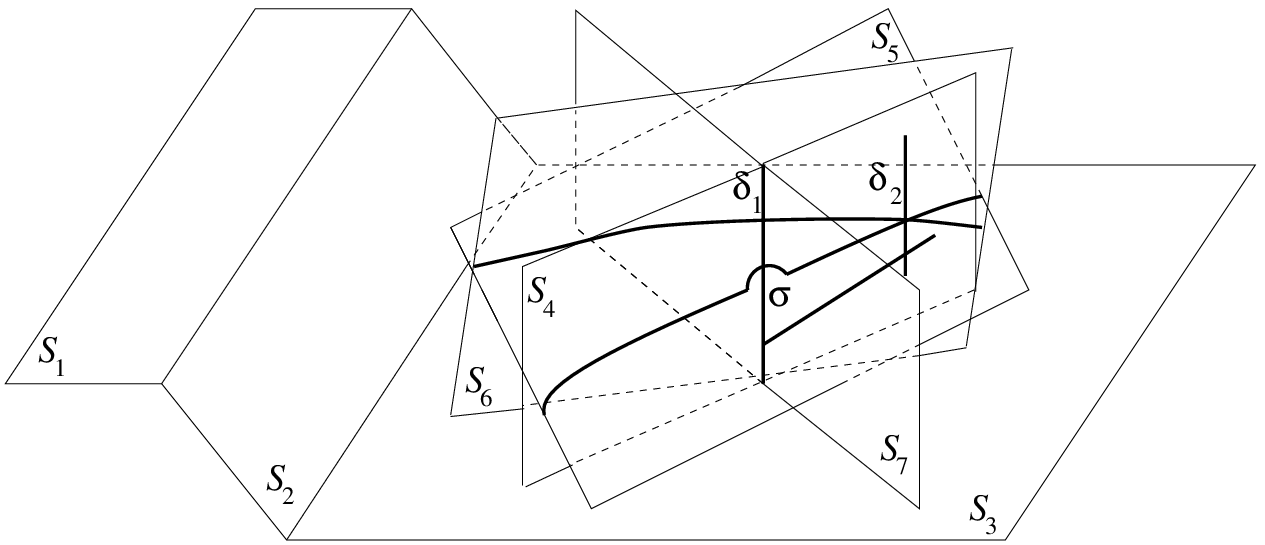}}

\noindent
All surfaces except $S_4$ are minimally ruled over rational curves. 
The base of $S_4$ is also rational but the surface is obtained from 
a minimal ruling by blowing-up two distinct points on a given fiber. 
The result is a reducible fiber with three components 
\eqn\redfib{
\epsilon_4=\delta_1+2\sigma+\delta_2.}
Note that $S_4$ and $S_7$ intersect along the curve $\delta_1$ which
is a fiber in the ruling of $S_7$. Similarly, $S_4$ and $S_6$
intersect along the curve $\delta_2$ which is a fiber in the ruling 
of $S_6$. This is consistent with the embedding in the ambient
Calabi-Yau space only if $\delta_1,\delta_2$ are $(-2)$ curves 
and $\sigma$ is a $(-1)$ curve on $S_4$. A local computation 
(appendix A) shows that this is 
indeed the case. The intersection matrix of 
$\delta_1,\delta_2$ and $\sigma$
with the exceptional divisors is 
\eqn\intmatB{\vbox{
\offinterlineskip
\halign{\strut
\quad $#$\quad &\vrule\hfill\quad $#$\quad\hfill&\vrule\quad 
$#$\quad\hfill&\vrule\hfill\quad $#$\quad\hfill&\vrule\hfill\quad 
$#$\quad\hfill&\vrule\hfill\quad $#$\quad\hfill\cr
&S_3&S_4&S_5&S_6&S_7\cr
\noalign{\hrule}
\delta_1&1&0&0&0&-2\cr
\noalign{\hrule}
\delta_2&0&0&1&-2&0\cr
\noalign{\hrule}
\sigma&0&-1&0&1&1\cr
}}}
From \redfib\ and \intmatB\ it follows that $\epsilon_i\cdot S_j$ 
reproduces the $E_7$ Cartan matrix. At the same time, 
$\sigma$ is a weight vector of the ${\bf 56}$
representation and all other vectors can be obtained by taking linear
combinations with the fiber classes $\epsilon_i$. 

The $\bf{56}$ representation of $E_7$ is pseudo-real, inducing a split of
its weight vectors into two sets $\Delta_\pm$, such that if
$v\in\Delta_+$, $(-\!v)\in\Delta_-$. Hypermultiplets
coming from anti-membranes wrapping curves associated to weight
vectors in $\Delta_-$ are therefore related to those arising from membranes
wrapping curves in $\Delta_+$, so a single holomorphic curve gives
rise to a half-hypermultiplet. Since $\pi_4(E_7)$ is trivial, the
$E_7$ gauge theory is not afflicted with a global anomaly, permitting
an odd number of half-hypermultiplets. This should be reflected in the
geometry, and indeed the weight $\sigma$ need not appear in
pairs. It is also consistent with the Weierstrass model considered
in appendix C.

Theories with  $n$
half-hypermultiplets can be engineered by colliding the surfaces $S_4$
and $S_7$ $n$ times along distinct fibers  $\delta_1^1,\ldots,\delta_1^n$ in
the ruling of $S_7$. This introduces $n$ reducible fibers 
\eqn\redfibB{
\epsilon^\alpha=\delta_1^\alpha+2\sigma^\alpha+\delta_2^\alpha,
\qquad \alpha=1,\ldots, n}
in the ruling of $S_4$. 

As in the $E_6$ case, the degeneration described in \Esevenone\ is not
unique and the relative \ka\ cone admits a subdivision into sub-cones
corresponding to different geometric phases. To see that note that
the present sub-cone, labeled phase I, is defined as the intersection
of the extended cone 
\eqn\extcone{
-S\cdot \epsilon_i>0}
with the hyper-cone
\eqn\hypconA{
-S\cdot\sigma>0}
which gives 
\eqn\sevenone{
\phi_6+\phi_7<\phi_4.}

We now proceed to uncover the phase structure as in the $E_6$
case. Since the $E_7$ Dynkin diagram does not have the
symmetry of the $E_6$ one, we will have to enumerate all the phases
explicitly.

${\underline{\hbox{Phase II}}}$ - The degeneration of this phase is
described in fig. 9 below.
\ifig\Eseventwo{$E_7$ degeneration - phase II}
{\epsfxsize4.0in\epsfbox{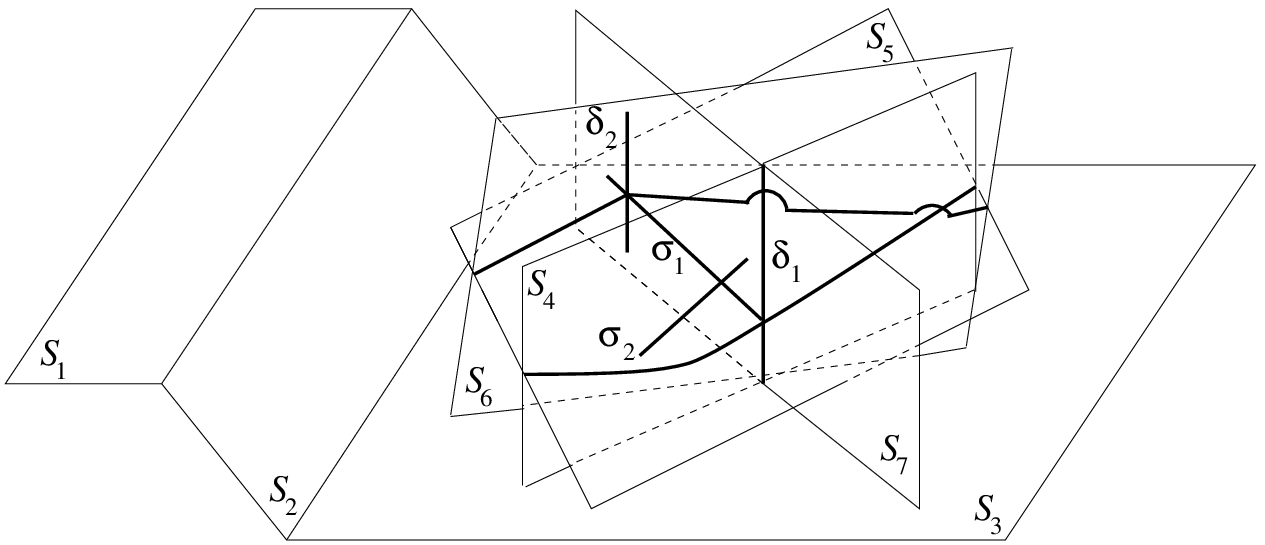}}

\noindent
All surfaces are
minimally ruled except for $S_5$ and $S_7$. $S_7$ has a reducible
fiber with three components
\eqn\redfibC{
\epsilon_7=\delta_1+2\sigma_1+\delta_2}
where $\delta_1,\delta_2$ are $(-2)$ curves and $\sigma_1$ is a $(-1)$
curve. $S_5$ has a reducible fiber with two components
\eqn\redfibD{
\epsilon_5=\sigma_1+\sigma_2}
where both $\sigma_1$ and $\sigma_2$ are $(-1)$ curves. 
Note that $S_4$ and $S_7$ intersect along $\delta_1$ which is a fiber
in the ruling of $S_4$. $S_5$ and $S_7$ 
intersect along $\sigma_1$ and $S_6$ and $S_7$ intersect along
$\delta_2$ which is a fiber in the ruling of $S_6$. The intersection 
matrix of $\delta_1,\delta_2,\sigma_1$ and $\sigma_2$
with the exceptional divisors is 
\eqn\intmatC{
\vbox{
\offinterlineskip
\halign{\strut
\quad $#$\quad &\vrule\hfill\quad $#$\quad\hfill&\vrule\quad 
$#$\quad\hfill&\vrule\hfill\quad $#$\quad\hfill&\vrule\hfill\quad 
$#$\quad\hfill&\vrule\hfill\quad $#$\quad\hfill\cr
&S_3&S_4&S_5&S_6&S_7\cr
\noalign{\hrule}
\delta_1&1&-2&1&0&0\cr
\noalign{\hrule}
\delta_2&0&0&1&-2&0\cr
\noalign{\hrule}
\sigma_1&0&1&-1&1&-1\cr
\noalign{\hrule}
\sigma_2&0&0&-1&0&1\cr
}}}
It is easily verified that \redfibD\ and \intmatC\ yield the correct 
Cartan matrix and that $\sigma_1,\sigma_2$ are weight vectors of 
${\bf 56}$. The inequalities which define this sub-cone can be read
off from \intmatC\ and \hypconA,
\eqn\seventwo{\eqalign{
\phi_4+\phi_6&<\phi_5+\phi_7,\cr
\phi_7&<\phi_5.}}
By adding $\phi_6$ to first line above and using the inequality 
$2\phi_6-\phi_5>0$ which holds everywhere
in the external cone we get $\phi_4<\phi_6+\phi_7$. This shows that 
phases I and II are indeed disjoint.

${\underline{\hbox{Phase III}}}$ -
The present phase can be immediately obtained from the previous one by
flopping $\sigma_5$ from $S_5$ to $S_7$. This gives the cone 
\eqn\seventhree{\eqalign{\phi_4+\phi_6&<\phi_5+\phi_7,\cr
\phi_5&<\phi_7.}}
The first inequality above is actually redundant. This can be seen by 
adding $\phi_5$ to both sides of the second inequality and using 
$2\phi_5-\phi_4-\phi_6>0$ which defines the external cone to get 
the first inequality in \seventhree. We can therefore
show, as in phase II, that phases I and III are disjoint.

So far, we have constructed three disjoint phases. The complement of
these three sub-cones in the extended \ka\ cone is defined by 
\eqn\compl{\eqalign{
\phi_4&<\phi_6+\phi_7,\cr
\phi_5+\phi_7&<\phi_4+\phi_6,}}
and we will have to verify that the inequalities defining the phases
below cover this region completely.

${\underline{\hbox{Phase IV}}}$ -
The degeneration corresponding to this phase is described in the figure
below.
\ifig\Esevenfour{$E_7$ degeneration - phase IV}
{\epsfxsize4.0in\epsfbox{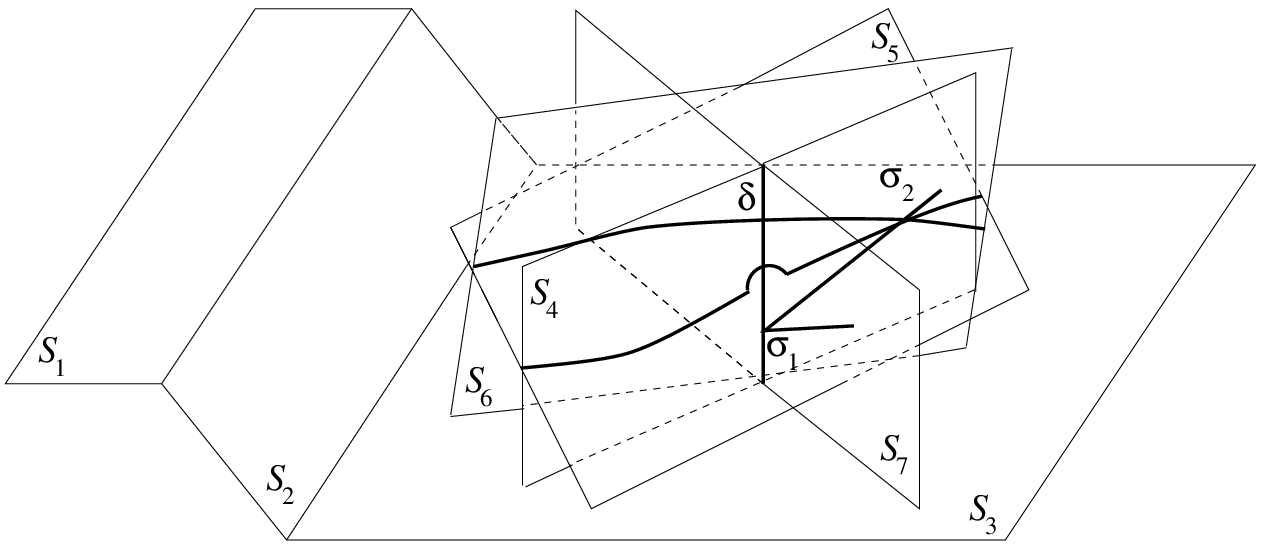}}
\noindent
In this sub-cone, $S_4$, $S_6$ and $S_7$ are minimally 
ruled surfaces blown-up in $n$ distinct points. The $n$ reducible 
fibers are 
\eqn\redfibG{\eqalign{
\epsilon_4^\alpha=\delta^\alpha+\sigma_2^\alpha\cr
\epsilon_6^\alpha=\sigma_1^\alpha+\sigma_2^\alpha\cr
\epsilon_7^\alpha=\delta^\alpha+\sigma_1^\alpha.\cr}}
Note that 
\eqn\dbintB{
S_4\cap S_7\simeq \sum_{\alpha=1}^n\delta^\alpha,\qquad
S_4\cap S_6\simeq \sum_{\alpha=1}^n\sigma_2^\alpha,\qquad
S_6\cap S_7\simeq \sum_{\alpha=1}^n\sigma_1^\alpha.}
The intersection matrix of $\delta,\sigma_1$ and $\sigma_2$ is 
\eqn\intmatB{\vbox{
\offinterlineskip
\halign{\strut
\quad $#$\quad &\vrule\hfill\quad $#$\quad\hfill&\vrule\quad 
$#$\quad\hfill&\vrule\hfill\quad $#$\quad\hfill&\vrule\hfil
\quad $#$\quad\hfill&\vrule\hfill\quad $#$\quad\hfill\cr
&S_3&S_4&S_5&S_6&S_7\cr
\noalign{\hrule}
\delta&1&-1&0&1&-1\cr
\noalign{\hrule}
\sigma_1&0&1&0&-1&-1\cr
\noalign{\hrule}
\sigma_2&0&-1&1&-1&1\cr
}}}
Thus the inequalities defining this phase are
\eqn\sevenfour{\eqalign{\phi_3+\phi_6&<\phi_4+\phi_7,\cr
\phi_4&<\phi_6+\phi_7,\cr
\phi_5+\phi_7&<\ph_4+\phi_6.}}

${\underline{\hbox{Phase V}}}$ -
The degeneration is described fig. 11 below.
\ifig\Esevenfive{$E_7$ degeneration - phase
V}{\epsfxsize4.0in\epsfbox{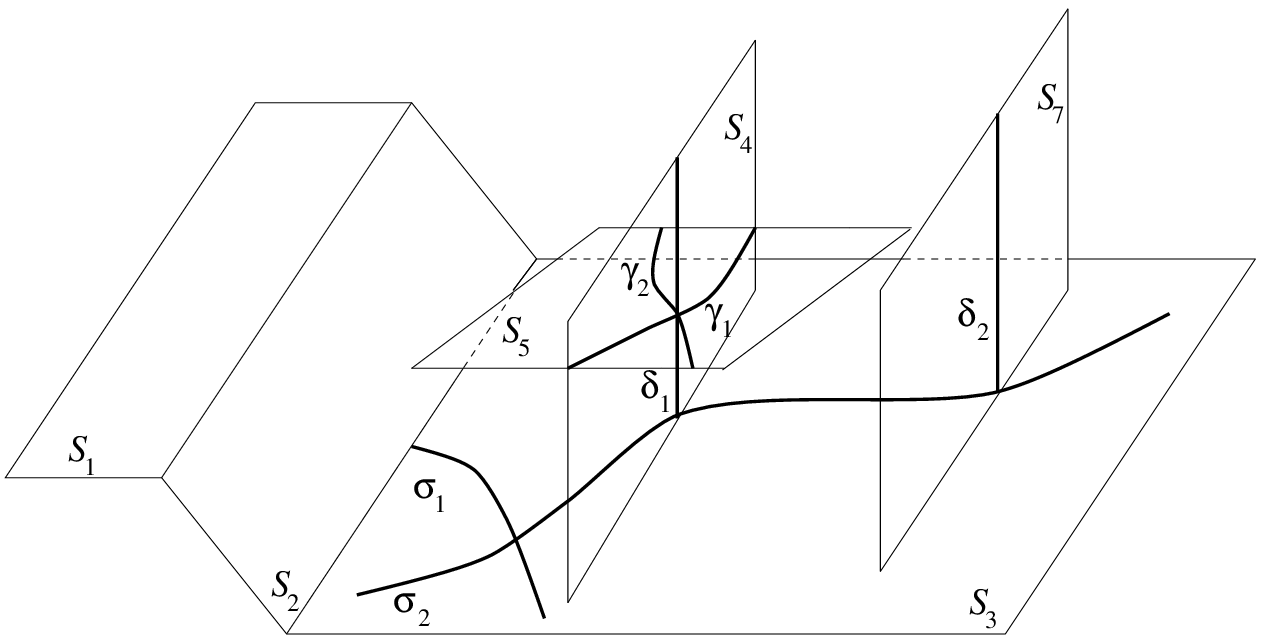}}
\noindent
All surfaces are minimally ruled except for $S_3$ and $S_6$ which is
too complicated to draw, but intersects $S_5$ along $\gamma_2$. The
reducible fibers are given by 
\eqn\fiberV{\eqalign{\epsilon_3&=\sigma_1+\sigma_2\cr
\epsilon_6&=\delta_1+\delta_2+2\sigma_2.}}
$\sigma_1$ is a $(-1)$ curve in $S_3$ and $\sigma_2$ is a $(-1)$
curves in both $S_3$ and $S_6$. $\delta_1$ and $\delta_2$ are $(-2)$
curves in $S_6$. The intersection matrix is given by
\eqn\intmatD{\vbox{
\offinterlineskip
\halign{\strut
\quad $#$\quad &\vrule\hfill\quad $#$\quad\hfill&\vrule\quad 
$#$\quad\hfill&\vrule\hfill\quad $#$\quad\hfill&\vrule\hfill\quad
$#$\quad\hfill&\vrule\hfill\quad 
$#$\quad\hfill&\vrule\hfill\quad $#$\quad\hfill\cr
&S_2&S_3&S_4&S_5&S_6&S_7\cr
\noalign{\hrule}
\delta_1&0&1&-2&1&0&0\cr
\noalign{\hrule}
\delta_2&0&1&0&0&0&-2\cr
\noalign{\hrule}
\sigma_1&1&-1&0&0&1&0\cr
\noalign{\hrule}
\sigma_2&0&-1&1&0&-1&1\cr
}},}
from which we can read the inequalities
\eqn\sevenfive{\eqalign{
\phi_4+\phi_7&<\phi_3+\phi_6\cr
\phi_2+\phi_6&<\phi_3.}}
It is easy to verify that this sub-cone is inside the complement of
phases I, II and III. Adding $\phi_7$ to both sides of the first
inequality in \sevenfive\ and using the $\epsilon_7$ inequality in
\extcone\ leads 
to the first inequality in \compl. Doing the same with $7\rightarrow4$
leads to the second inequality in \compl.

${\underline{\hbox{Phases VI, VII and VIII}}}$ -
These phases are obtained by flopping $\sigma_2$ first to $S_2$ then
to $S_1$ and finally outside $S_1$.

\ifig\Esevensix{$E_7$ degenerations - phases VI, VII and VIII}
{\epsfxsize5.4in\epsfbox{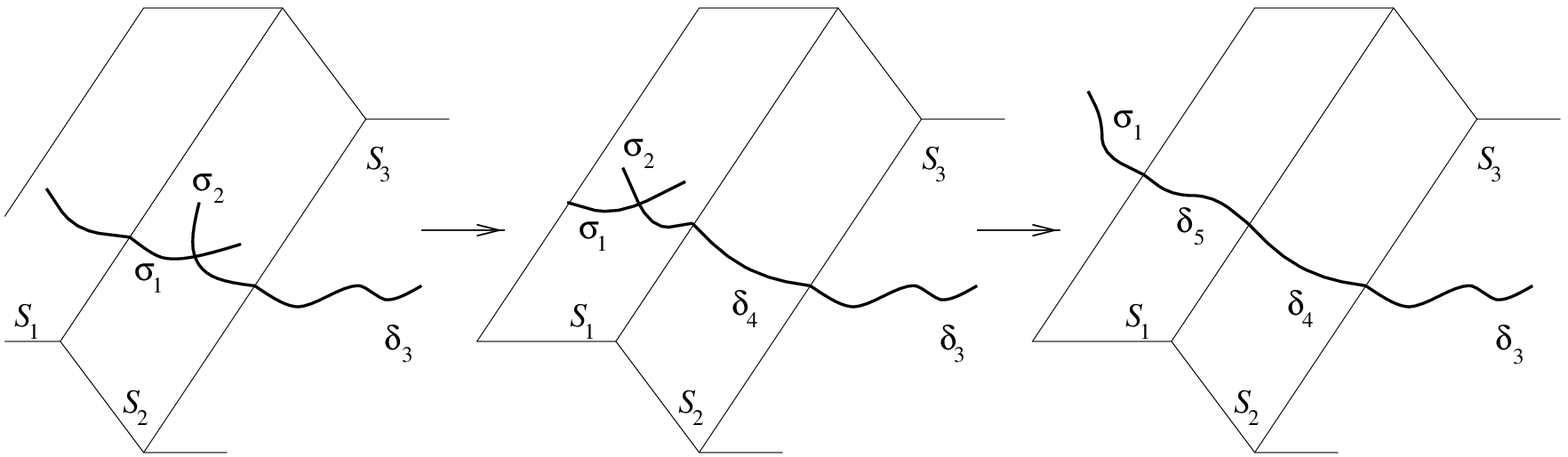}}

\noindent
The geometry of these degenerations is identical to that of phase V
with the following differences:
\eqn\diffseven{\vbox{
\offinterlineskip
\halign{\strut
\quad $#$\quad&\vrule\quad $#$\quad\hfill&\vrule\quad 
$#$\quad\hfill&\vrule\quad $#$\quad\hfill\cr
&\hbox{phase VI}&\hbox{phase VII}&\hbox{phase VIII}\cr
\noalign{\hrule}
\epsilon_1&\epsilon_1&\sigma_1+\sigma_2&\delta_5\cr
\noalign{\hrule}
\epsilon_2&\sigma_1+\sigma_2&\delta_4&\delta_4\cr
\noalign{\hrule}
\epsilon_3&\delta_3&\delta_3&\delta_3\cr
\noalign{\hrule}
\epsilon_6&{\eqalign{&\delta_1+\delta_2+2\delta_3\cr&+2\sigma_2}}
&{\eqalign{&\delta_1+\delta_2+2\delta_3\cr&+2\delta_4+2\sigma_2}}
&{\eqalign{&\delta_1+\delta_2+2\delta_3\cr&+2\delta_4+2\delta_5}}\cr
}}}
It straight forward  to check how the intersection numbers of the
curves $\sigma_{1,2}$ change as the double fiber is flopped from phase
V to phase VIII. We find the following inequalities
\eqn\sevensix{\eqalign{\hbox{phase VI}&\qquad\phi_3<\phi_2+\phi_6\cr
&\qquad\phi_1+\phi_6<\phi_2\cr
\hbox{phase VII}&\qquad\phi_2<\phi_1+\phi_6\cr
&\qquad\phi_6<\phi_1\cr
\hbox{phase VIII}&\qquad\phi_1<\phi_6.}}
The calculation showing that all these phases are mutually disjoint as
well as disjoint from the previous ones is easy and will not be
repeated here.

To summarize, in the presence of massless ${\bf 56}$ matter
multiplets, the $E_7$ Coulomb branch is divided into eight phases
determined by the inequalities \sevenone, \seventwo, \seventhree, 
\sevenfour, \sevenfive\ and \sevensix.

\subsec{A prepotential calculation}

We calculate the prepotential in phase IV defined in \sevenfour.
The triple intersections of this configuration 
can be computed as follows. $S_4$ and $S_7$ intersect $S_3$ along two 
sections
meeting transversely $n$ times. As $S_3$ is minimally ruled, it
follows that it must be isomorphic to $\f_n$ and the two sections 
are in the $C^{\infty}_3$ class. $S_2$ does not intersect $S_4$ and 
$S_7$, therefore it must meet $S_3$ along $C^0_3$. This results in the
intersection numbers
\eqn\intEseven{\eqalign{
& S_3S_4^2=S_3S_7^2=n,\qquad\hskip 25pt S_3S_2^2=-n\cr
& S_3^2S_4=S_3^2S_7=-n-2,\qquad S_3^2S_2=n-2.\cr}}
It is then easy to compute
\eqn\intnoG{
S_1^2S_2=2-n,\qquad S_1S_2^2=n-4.}
Since $S_1$ and $S_2$ intersect along $C^{\infty}_1$ on $S_1$ and 
$C^0_2$ on $S_2$, the degrees of the rulings are 
\eqn\degE{
n_1=n-4,\qquad n_2=n-2,\qquad n_4=n_7=n+2.}
Note that the non-minimal surfaces $S_4,S_6$ and $S_7$ are obtained from
minimal rulings by blowing-up the intersection points of $n$ fibers 
with a section in $C^\infty$. 
Therefore, $S_4$ and $S_5$ intersect along the strict transform
$C^{\infty}_4-\sum_{\alpha=1}^n\sigma_2^\alpha$ on $S_4$ and $C^0_5$ on 
$S_5$. It follows that 
\eqn\intnoH{
S_4S_5^2=2,\qquad S_4^2S_5=-4,\qquad n_5=4.}
The intersections \dbintB\ determine
\eqn\intnoI{\eqalign{
& S_4^2S_7=S_4S_7^2=-n\qquad\qquad S_3S_4S_7=n\cr
& S_4^2S_6=S_4S_6^2=-n\qquad\qquad S_4S_6S_7=n\cr
& S_6^2S_7=S_6S_7^2=-n\qquad\qquad S_4S_5S_6=n.}}
$S_4$ and $S_6$ intersect $S_5$ along two sections meeting in $n$
points lying on the curves $\sigma_2^\alpha$. Since $n_5=4$, the two
sections must be $C^0_5$ and $C^{\infty}_5+n\epsilon_5$ respectively. 
This implies that 
\eqn\intnoK{
S_5S_6^2=2n+4,\qquad S_5^2S_6=-2n-6,\qquad n_6=2n+6.}
Finally, the remaining intersection numbers are 
\eqn\intnoJ{
S_1^3=S_2^3=S_3^3=S_5^2=8,\qquad S_4^3=S_6^3=S_7^3=8-n.}
The intersection numbers \intEseven-\intnoJ\ lead to the prepotential
\eqn\prepeseven{\eqalign{{\cal{F}}&=
8\phi_1^3-3(n-2)\phi_1^2\phi_2+3(n-4)\phi_1\phi_2^2 +
8\phi_2^3-3n\phi_2^2\phi_3+3(n-2)\phi_2\phi_3^2+8\phi_3^3\cr
&-3(n+2)\phi_3^2\phi_4+3n\phi_3\phi_4^2+
(8-n)\phi_4^3-12\phi_4^2\phi_5+6\phi_4\phi_5^2+
8\phi_5^3-3n\phi_4^2\phi_6+6n\phi_4\phi_5\phi_6\cr
&-3(2n+6)\phi_5^2\phi_6-3n\phi_4\phi_6^2+
3(2n+4)\phi_5\phi_6^2+(8-n)\phi_6^3-
3(n+2)\phi_3^2\phi_7+6n\phi_3\phi_4\phi_7\cr
&-3n\phi_4^2\phi_7+6n\phi_4\phi_6\phi_7-
3n\phi_6^2\phi_7+3n\phi_3\phi_7^2-3n\phi_4\phi_7^2-
3n\phi_6\phi_7^2+(8-n)\phi_7^3}}
This agrees with the gauge theory computation for this phase.

\subsec{Fixed points analysis}

The fixed point analysis follows the same lines as in the $E_6$ case
and we shall do it in phase IV. 
Let 
\eqn\gendivC{
S=\sum_{i=1}^7\phi_iS_i}
denote a generic divisor in the sub-cone \sevenfour.
The restrictions of $S$ to the surfaces $S_i$ are 
\eqn\dbintD{\eqalign{
H_1\equiv -S\cdot S_1=&(2\phi_1-\phi_2)C^{\infty}_1+
\phi_1(6-n)\epsilon_1\cr
H_2\equiv -S\cdot S_2=&(2\phi_2-\phi_1-\phi_3)C^{\infty}_2+
(4\phi_2-2\phi_1-n(\phi_2-\phi_1))\epsilon_2\cr
H_3\equiv -S\cdot S_3=& (2\phi_3-\phi_2-\phi_4-\phi_7)C^{\infty}_3+
(2\phi_3-n(\phi_3-\phi_2))\epsilon_3\cr
H_4\equiv -S\cdot S_4=& (2\phi_4-\phi_3-\phi_5)C^{\infty}_4+
(2\phi_3-n(\phi_4+\phi_7-\phi_3))\epsilon_4\cr
&-(\phi_4+\phi_6-\phi_5-\phi_7)\sum_{\alpha=1}^n\sigma_2^{\alpha}\cr
H_5\equiv -S\cdot S_5=&(2\phi_5-\phi_4-\phi_6)C^{\infty}_5+
(4\phi_4-2\phi_5-n\phi_6)\epsilon_5\cr
H_6\equiv-S\cdot
S_6=&(2\phi_6-\phi_5)C^{\infty}_6+(6\phi_5-4\phi_6-
n(2\phi_6-2\phi_5+\phi_4))\epsilon_6\cr
&-(\phi_6+\phi_7-\phi_4)\sum_{\alpha=1}^n\sigma_1^\alpha\cr
H_7\equiv-S\cdot S_7=&(2\phi_7-\phi_3)C^{\infty}_7+
(2\phi_3-n(\phi_7-\phi_3+\phi_4))\epsilon_7\cr
&-(\phi_6+\phi_7-\phi_4)\sum_{\alpha=1}^n\sigma_1^\alpha.
}}
We first check ampleness on the minimally ruled surfaces $S_1,S_2,S_3$
and $S_5$. Taking into account the inequalities \sevenfour\ defining
the phase, this yields the bound $n\leq 5$. Note that all surfaces allow
$n=6$ except for $S_1$. 

Next, we consider the blown-up surface $S_4$. As in the previous
section, a generic divisor class can be written
\eqn\gendivfour{
D_4=a_4C^0_4+b_4\epsilon_4+\sum_{\alpha=1}^nc_\alpha\sigma_2^\alpha.}
If this class admits an effective representative, it must have
non-negative intersection number with all irreducible holomorphic
curves $C$ such that $C^2\geq 0$. Checking that this condition holds
for the curves 
\eqn\curvesthree{
C^\infty_4, \quad \epsilon_4,\quad C^\infty_4-
\sum_{\alpha=1}^n\sigma_2^\alpha}
gives the restrictions
\eqn\restr{\eqalign{
&a_4\geq 0\cr
&b_4\geq 0\cr
&b_4+\sum_{\alpha=1}^n c_\alpha\geq0.}}
The intersection numbers of $H_4$ with $C_4^0,\sigma_4$ and $\delta$
are easily seen to be positive inside the sub-cone \sevenfour\ when
$n\leq 6$. Using the conditions \restr\ we then find that $H_4$ is
restricts to an ample class on $S_4$ if $n\leq 6$. The analysis of
$H_6$ and $H_7$ is identical and gives the same restriction on $n$.

In conclusion, we find that $H_i$ is ample on each surface $S_i$ if
$n\leq 5$. Therefore for $n_{56}={n\over 2}\leq {5\over{2}}$ there exist UV
fixed points with $E_7$ gauge symmetry.

\centerline{\bf Acknowledgments}

We would like to thank Kenneth Intriligator, Ori Ganor, Barak Kol,
Lubo\v{s} Motl, David Morrison, Christian R\"omelsberger, Nathan
Seiberg and especially Cumrun Vafa for helpful discussions. We are
very grateful to Cumrun Vafa for suggesting the idea of this work.

\appendix{A}{Ruled Surfaces, Blow-ups and All That}

This is a brief exposition of basic properties of ruled 
surfaces and their monoidal transformations used throughout
the paper. These facts can be found in any mathematical monograph 
on the subject. Our presentation follows \Ha.

A ruled surface is a smooth complex projective surface $S$ with 
a fibration structure 
$\pi:S\rightarrow C$ where $C$ is a smooth curve of genus $g$
and the fibers of $\pi$ are isomorphic to $\p^1$. In general, 
$S$ can be represented as the projectivization of a rank two 
holomorphic vector bundle $E$ on $C$. The degree $n=-degE$ is 
an invariant of $S$. 

The Picard group of $S$ contains two distinguished section 
classes, $C_0$ and $C_{\infty}$,
and a fiber class $f$. They satisfy
\eqn\Picrel{\eqalign{
& C_0^2=-n,\qquad f^2=0,\qquad C_0\cdot f=1\cr
& C_\infty=C_0+nf\quad \Rightarrow\quad C_\infty\cdot C_0=0,
\qquad C_\infty\cdot f=1.\cr}}
Note that the two sections are disjoint and 
\eqn\intrelAA{
C_0^2+C_\infty^2=0.}
The canonical class of $S$ is given by 
\eqn\cancls{
K_S=-2C_0-(n+2)f=-2C_\infty +(n-2)f.}

Other surfaces occuring in many examples in the main text 
are simply obtained by blowing-up the points 
$P_1,\ldots, P_k\in S$ lying at the the intersection of $k$ fibers with
a given section $\gamma$. The latter can be either $C_0$ or
$C_\infty$. For concreteness, we  consider $\gamma\simeq C_\infty$. 
The resulting surface, ${\tilde S}_1$, can be 
regarded as 
fibration over $C$ with a reducible fiber with two components
\eqn\redfibAA{
f=e_1+e_1^\prime =\ldots =e_k+e_k^\prime}
over $P_1,\ldots, P_k$ as in fig. 13. 

\ifig\exO{Blowing-up $k$ points lying on $C_\infty$.}
{\epsfxsize3.5in\epsfbox{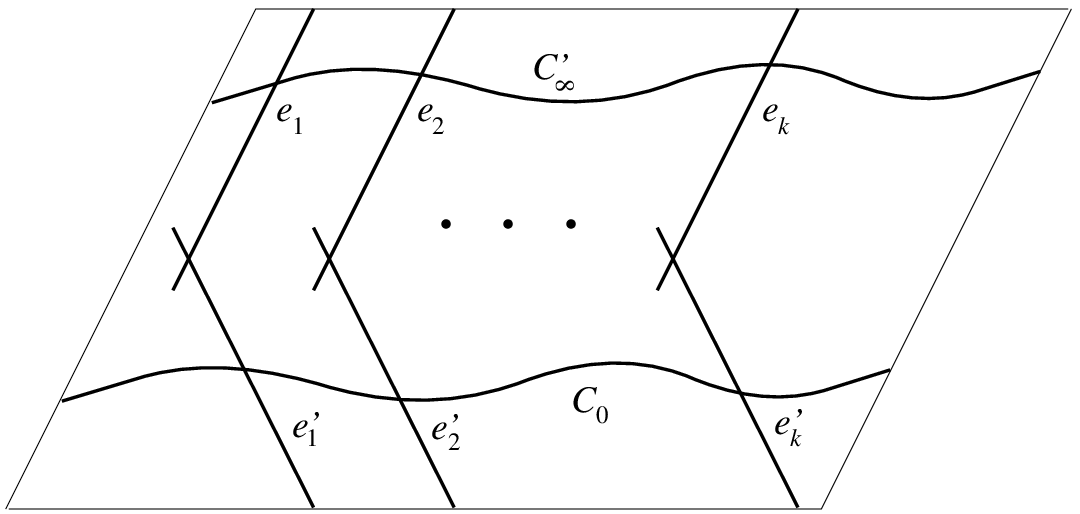}}

\noindent
The fiber components satisfy the intersection relations
\eqn\intrelAB{\eqalign{
& e_i\cdot e_j = e_i^\prime \cdot e_j^\prime =-\delta_{ij},
\qquad e_i\cdot e_j^\prime=
\delta_{ij}\cr
& e_i\cdot C_0 =e_i\cdot C_\infty =0,\qquad 
  e_i^\prime\cdot C_0 =1.}}
The strict transform of $C_\infty$ is 
\eqn\newcls{
C_\infty^\prime =C_\infty-\sum_{i=1}^k e_i,}
satisfying
\eqn\newrel{
e_i\cdot C_\infty^\prime =1,\qquad
e_i^\prime\cdot C_\infty^\prime =0.}
Note also that 
\eqn\intrelAC{
C_0^2+{C_\infty^\prime}^2=-k.}
The canonical class of ${\tilde S}_1$ is given by 
\eqn\canclsB{
K_{{\tilde S}_1}=-2C_0-(n+2)f+\sum_{i=1}^k e_i.}

This example can be generalized in a number of ways, which 
are important for the degenerations in section three and four. 
We can blow-up
a pair of points of intersection of each of the above $k$ fibers with 
two distinct sections. The latter can be again chosen to be
$C_0,C_\infty$ or two sections in $C_\infty$ (corresponding
respectively to surfaces $S_1$ and $S_4$ in section 3.2).

\ifig\exP{Blowing-up $2k$ points lying on two $C_\infty$ sections.}
{\epsfxsize3.5in\epsfbox{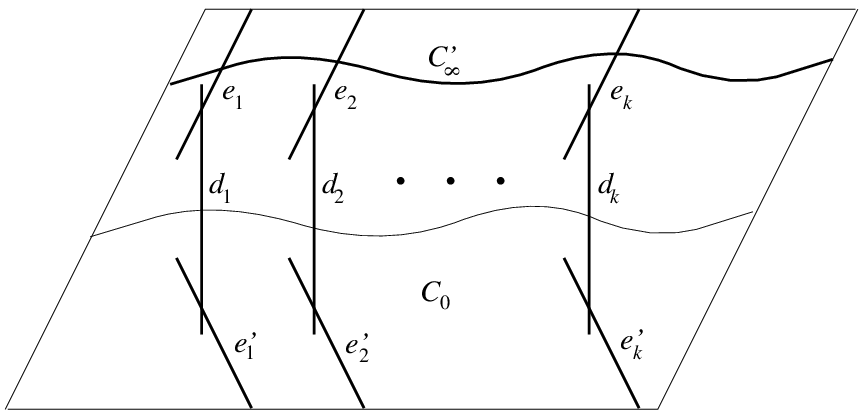}}

\noindent
This results in $k$ reducible fibers with three components
\eqn\redfibAB{
f=e_1+d_1+e_1^\prime =\ldots = e_k+d_k+e_k^\prime.}
The new intersection numbers are 
\eqn\newint{\eqalign{
& e_i\cdot e_j = e_i^\prime \cdot e_j^\prime =-\delta_{ij},\cr
& d_i\cdot d_j = -2\delta_{ij},\qquad e_i\cdot d_j=e_i^\prime\cdot 
d_j=\delta_{ij}.\cr}}
Note that the strict transforms of the original fibers $d_i$ 
must now be $(-2)$ curves. The original sections $C_0$ and $C_\infty$
do not meet the exceptional components $e_i$ or $e_i^\prime$.
The canonical class of this surface  is now given by 
\eqn\canclsC{
K_{{\tilde S}_2}= -2C_0-(n+2)f+\sum_{i=1}^k 
\left(e_i +e_i^\prime\right).}
This construction is relevant for phase I of the $E_6$ degeneration. 
Phases III and V-VII involve reducible fibers 
of the form $f=e+d_1+\ldots d_k+e^\prime$ where 
$e,e^\prime$ are $(-1)$ curves and $d_i$ are $(-2)$ curves.
These can be realized by successive blow-ups.

Alternatively, we can blow-up the $k$ points of intersection of the 
fiber components in fig. 13. The resulting surface has $k$
reducible fibers with three components
\eqn\redfibAD{
f=e_1+2d_1+e_1^\prime=\ldots = e_k+2d_k+e_k^\prime.}
Here $e_i$ and $e_i^\prime$ are proper transforms of the original fiber 
components while $d_i$ are $(-1)$ curves induced by the blow-up. 
Their intersection matrix is given by 
\eqn\intrelAD{
\eqalign{
& e_i\cdot e_j = e_i^\prime \cdot e_j^\prime =-2\delta_{ij},\cr
& d_i\cdot d_j = -\delta_{ij},\qquad e_i\cdot d_j=e_i^\prime\cdot 
d_j=\delta_{ij}.\cr}}
Note that $d_i$ appear with multiplicity $2$ in the fiber so that 
$f^2=0$ is satisfied. The original sections $C_0$ and $C_{\infty}$ 
intersect $e_i^\prime$ but not the other fiber components.
Different sections passing through the fiber components can be defined
as in \newcls.\
The canonical class of the new surface ${\tilde S}_3$ is given by 
\eqn\canclsD{
K_{{\tilde S}_3}=-2C_0-(n+2)f+\sum_{i=1}^k\left(2d_i+e_i\right).}

Phases VI-VIII in section four require reducible fibers of the form
$f=d_1+d_2+2d_3+\ldots 2d_k+2e$ where $d_i$ are $(-2)$ curves and $e$
is a $(-1)$ curve. This can be realized by successive blow-ups of 
${\tilde S}_3$. 

\ifig\exQ{Reducible fiber for the $E_7$ phases V-VIII.}
{\epsfxsize3.5in\epsfbox{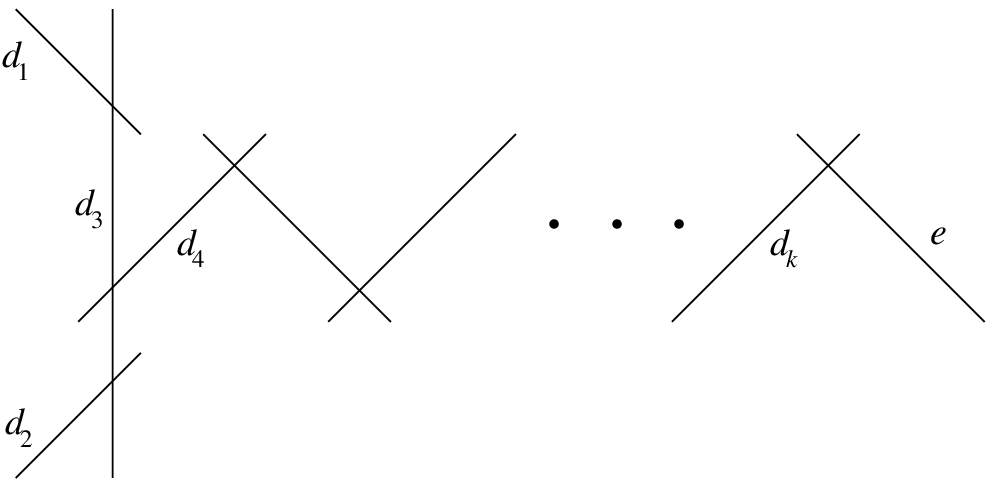}}

\noindent

We end this section with some comments on triple intersections in 
Calabi-Yau degenerations. The neighborhood of any surface $S$ embedded
in a smooth Calabi-Yau threefold $X$ is isomorphic to the total space of
the canonical bundle of $S$. Therefore the intersection numbers
are determined by local data
\eqn\CYintA{
S^3=K_S^2.}
This gives $S^3=8$ for a minimally ruled surface. Equations \canclsB,\
\canclsC\ and \canclsD\ show that $S^3$ decreases by $1$ for each
blow-up of $S$
\eqn\CYintB{
{\tilde S}_1^3=8-k,\qquad {\tilde S}_2^3={\tilde S}_3^3=8-2k.}
Next, if two surfaces $S_1,S_2$ intersect transversely
along a curve $\gamma$ we have 
\eqn\CYintC{
S_1^2S_2=(\gamma^2)_{S_2},\qquad S_1S_2^2=(\gamma^2)_{S_1}.}
The embedding in the Calabi-Yau space requires
\eqn\CYintD{
S_1^2S_2+S_1S_2^2=deg\left(N_{\gamma /S}\right)=2g-2}
where $g$ is the genus of $\gamma$. 
Combined with \intrelAC,\ these are the basic elements entering the 
prepotential computations in the main text. 

\appendix{B}{Resolution of the $E_6$ Weierstrass Model}

We consider six dimensional F theory compactified on a singular 
Weierstrass model over $\f_n$ base described by the equation
\eqn\WrA{
y^2=x^3+xf(z_1,z_2)+g(z_1,z_2).}
Here $z_1,z_2$ are affine coordinates on the $\p^1$ fiber and $\p^1$
base of the Hirzebruch surface. According to \BIK,\ the Weierstrass
model exhibits a split $E_6$ singularity along the section $z_1=0$ 
of $\f_n$ if 
\eqn\slitsingA{\eqalign{
& f(z_1,z_2)=z_1^3f_{8+n}(z_2)\cr
& g(z_1,z_2)=z_1^4q_{n+6}^2(z_2)+z_1^5g_{12+n}(z_2)\cr}}
where the subscripts indicate the degrees of the polynomials. 
The discriminant of the elliptic fibration is 
\eqn\discrA{
\Delta=z_1^8\left(27q_{n+6}(z_2)^4+O(z_1)\right).}
Therefore the line $z_1=0$ of $IV^*$ singularities intersects 
a line of $I_1$ singularities at the zeroes of $q_{n+6}$. 
Based on the dual heterotic model, each zero of $q_{n+6}$ 
is expected to yield a hypermultiplet in the ${\bf 27}$
representation. We will explicitly check this prediction by
constructing a smooth model of the singular elliptic fibration 
as in \refs{\RM,\A}. 

Introducing local coordinates $(t,s)$ near a simple 
zero of $q_{n+6}$,
the singularity can be written 
\eqn\singA{
y^2=x^3+s^3x+s^4t^2.}
In these coordinates, we have 
\eqn\singB{
f=s^3,\qquad g=s^4t^2,\qquad \Delta = s^8\left(4s+27t^4\right).}
The vanishing degrees along the line $s=0$ in the base are 
$(3,4,8)$ therefore this is a line of $IV^*$ fibers. 
The vanishing degrees along $4s+27t^4=0$ are $(0,0,1)$ therefore
this is a line of $I_1$ fibers colliding $s=0$ at $s=t=0$. 
At the collision locus, the degrees jump to $(3,6,9)$ which 
characterize a $III^*$ fiber. Although the collision is not 
transverse, we can still apply the resolution scheme of 
\RM\ (see the appendix of \refs{\A}). The elliptic fibration can be 
represented as a double cover of 
a $\p^1(x)$ fibration over the $(s,t)$ plane branched over the 
surface $B$
\eqn\branchA{
x^3+s^3x+s^4t^2=0.}

The problem can be reduced to the elliptic surface case by
considering certain slices through various points of the discriminant
\A. 
For fixed $t\neq 0$ this is a singularity of analytic type 
$x^3+s^4$ describing the generic $IV^*$ fiber. For $t=0$, the
singularity is of analytic type $x^3+s^3x$ corresponding to a 
$III^*$ fiber. The resolution proceeds by successively blowing up 
the $x=s=0$ line until the generic $IV^*$ singularity is resolved. 
The $III^*$ fiber over $s=t=0$ is only partially resolved after this
process. The intermediate steps are represented in fig. 16 and fig. 17.
According to the conventions of \refs{\RM,\A}, dotted line segments 
represent unbranched rational curves and solid line segments represent
branched exceptional rational curves. The curved lines
represent the proper transform of the original branch locus. 
The leftmost dotted vertical line always represents the proper
transform of the original fiber of the ruling. In the M theory limit,
this component grows to infinite size. The curves $f_1\dots f_4$ 
are fibers of the exceptional ruled surfaces $\f_1,\ldots,\f_4$ over 
the $t$-line. Note that the proper transform $\tilde B$ of the branch 
surface intersects $\f_1$ and $\f_2$ transversely along the fibers $\f_1$
and $\f_2$ over
$s=t=0$. The surface $\f_4$ is branched while $\f_3$ intersects the
branch locus along two sections. 

\ifig\exI{Resolution of the generic $IV^*$ fiber.}
{\epsfxsize3.7in\epsfbox{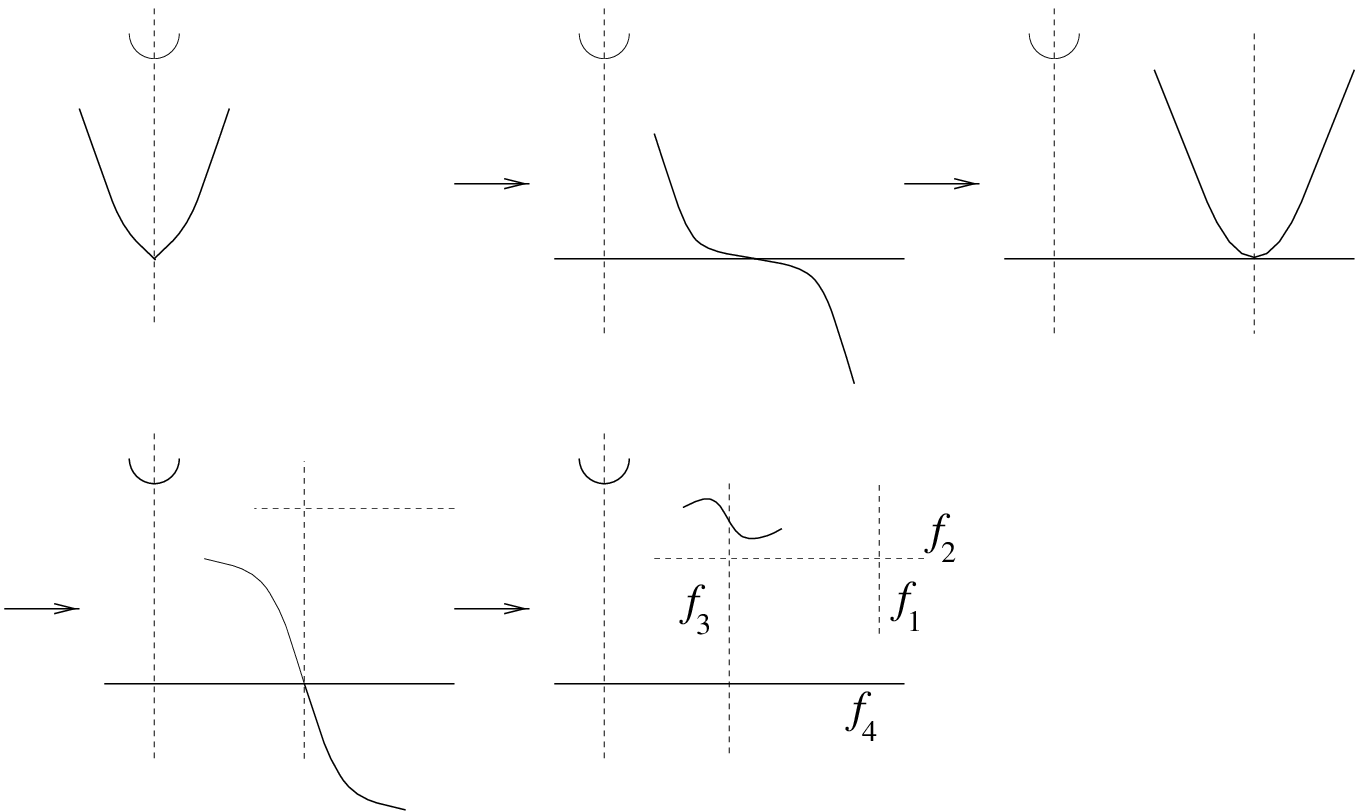}}

\ifig\exJ{Partial resolution of the $III^*$ fiber 
over the collision locus.}
{\epsfxsize3.7in\epsfbox{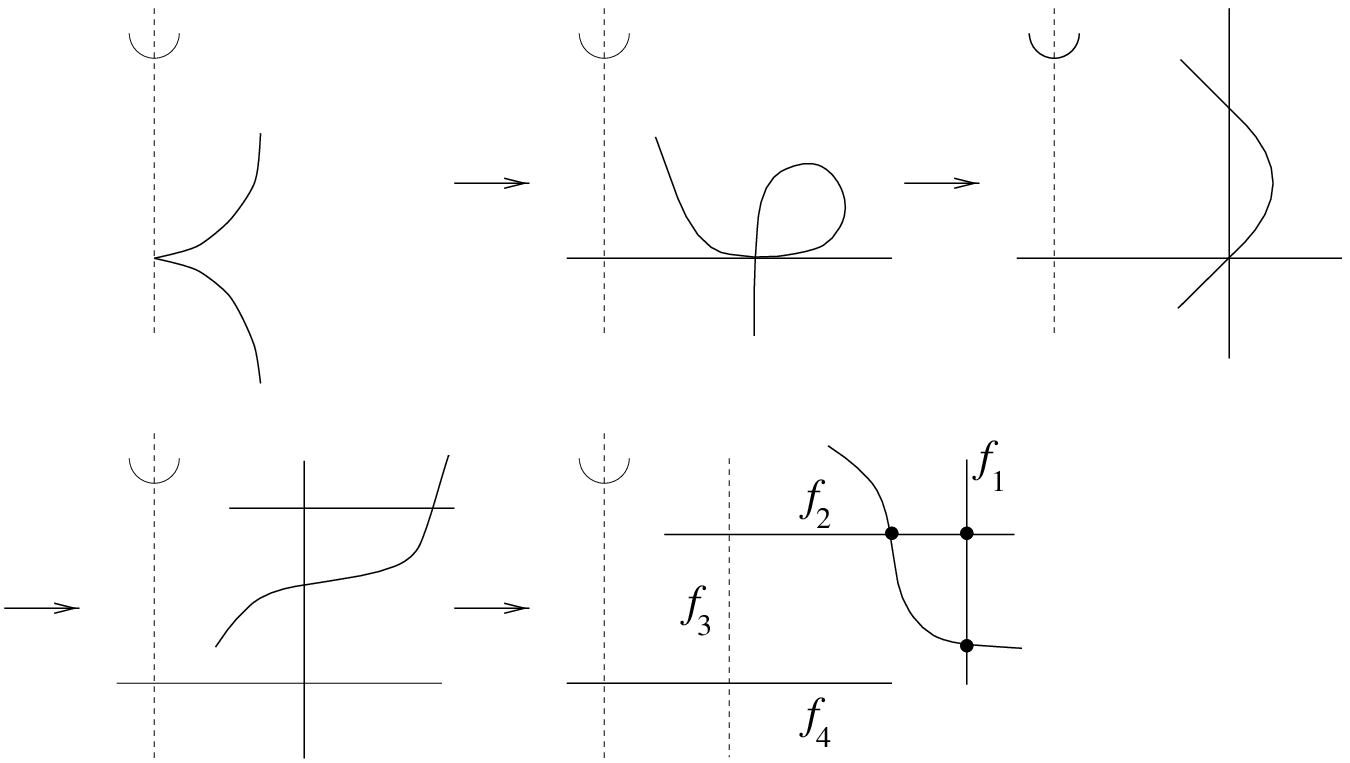}}

\noindent
The three marked points in the last diagram in fig. 17. represent 
three double point singularities of $\tilde B$ of analytic type 
$uv=w^2$ in certain coordinates $(u,v,w)$ on the blow-up. 
Therefore the threefold is still
singular at this point. In order to construct a smooth model, 
one has to blow-up the three singular points \RM\ introducing three 
exceptional $\p^2$ divisors as in fig. 18.a.

\ifig\exK{Construction of the smooth threefold model. The
configuration $b$ represents the double cover of $a$.}
{\epsfxsize3.7in\epsfbox{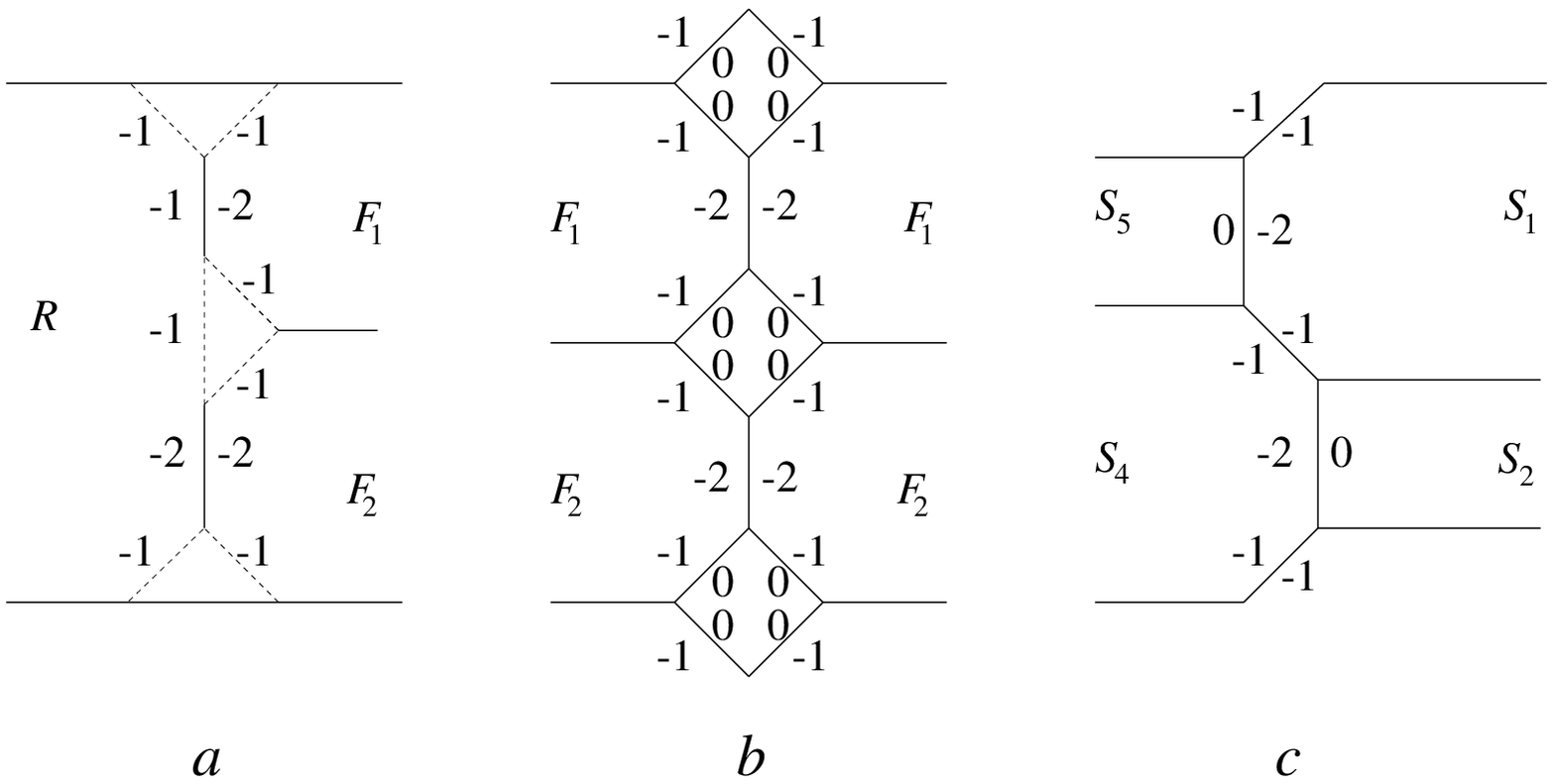}}

\noindent
The surface $R$ is the proper 
transform of the original $\p^1(x)$ ruling over the $s$-line $t=0$. 
The numbers indicate the bidegrees of the normal bundles of various 
rational curves. Taking double cover of the blown-up threefold yields the
configuration represented in fig. 18.b. The surface $R$ is no longer 
explicitly represented there.  Note that the exceptional 
$\p^2$ components are covered by $\f_0$ surfaces. Finally, a smooth 
threefold model with one-dimensional fibers can be obtained by
contracting the exceptional $\f_0$ surfaces in any direction. 
A possible contraction and the resulting smooth model are described
in fig. 18c. It can be easily checked that this is precisely the $E_6$
degeneration of section three. For clarity, we have identified the
exceptional ruled surfaces. Different contractions of the $\f_0$
surfaces result in different geometric phases as in section three. 

\appendix{C}{Resolution of the $E_7$ Weierstrass Model}

We consider a singular Weierstrass model with 
\eqn\WrB{\eqalign{
& f(z_1,z_2)=z_1^3f_{8+n}(z_2)\cr
& g(z_1,z_2)=z_1^5g_{12+n}(z_2).\cr}}
The discriminant of the elliptic fibration is given by 
\eqn\discrB{
\Delta=z_1^9\left(4f_{8+n}^3(z_2)+O(z_1)\right).}
Taking into account the vanishing orders of $f,g, \Delta$ \BIK,\ 
there is a line $z_1=0$ of $III^*$ fibers colliding a second line of 
$I_1$ singularities at the zeroes of $f_{8+n}$. Based on the dual
heterotic model, there should each zero of $f_{8+n}$ should correspond
to a half-hypermultiplet in the ${\bf 56}$ representation.

The smooth flat model can be constructed as follows. The singularity
can be written locally near a simple zero of $f_{8+n}$ as
\eqn\locsing{
y^2=x^3+s^3tx+s^5.}
We have 
\eqn\locsingB{
f=s^3t,\qquad g=s^5,\qquad \Delta = s^9(4t^3+27s).}
The vanishing degrees of $(f,g,\Delta)$ along $s=0$ are 
$(3,5,9)$ characterizing a $III^*$ fiber. The vanishing degrees 
over $4t^3+27s=0$ are $(0,0,1)$, therefore we obtain a line of $I_1$
fibers colliding $s=0$ at $s=t=0$. At the collision locus, the
vanishing degrees jump to $(4,5,10)$ signaling a $II^*$ fiber. 
This jump is expected to give rise to a half-hypermultiplet in 
the ${\bf 56}$ representation. We explicitly show that this is the 
case by constructing the corresponding smooth model. 

The elliptic fibration can be represented as a double cover 
of a $\p^1(x)$ bundle over the $(s,t)$ plane branched over the surface
$B$
\eqn\branchC{
x^3+s^3tx+s^5=0.}
As in the previous case, the problem can be reduced to the elliptic
surface case by taking slices through points in the discriminant. 
For fixed generic $t\neq 0$ the singularity is of analytic type 
$x^3+s^3x$ corresponding to a $III^*$ fiber over the $t$-line. 
For $t=0$, the singularity is of analytic type $x^3+s^5$ which
corresponds to a $II^*$ fiber. The resolution follows the 
steps outlined in appendix B. First, we successively blow-up the 
line $x=s=0$ until the generic $III^*$ fiber is resolved. This 
process is represented in fig. 19.

\ifig\exL{Resolution of the generic $III^*$ fiber.}
{\epsfxsize3.7in\epsfbox{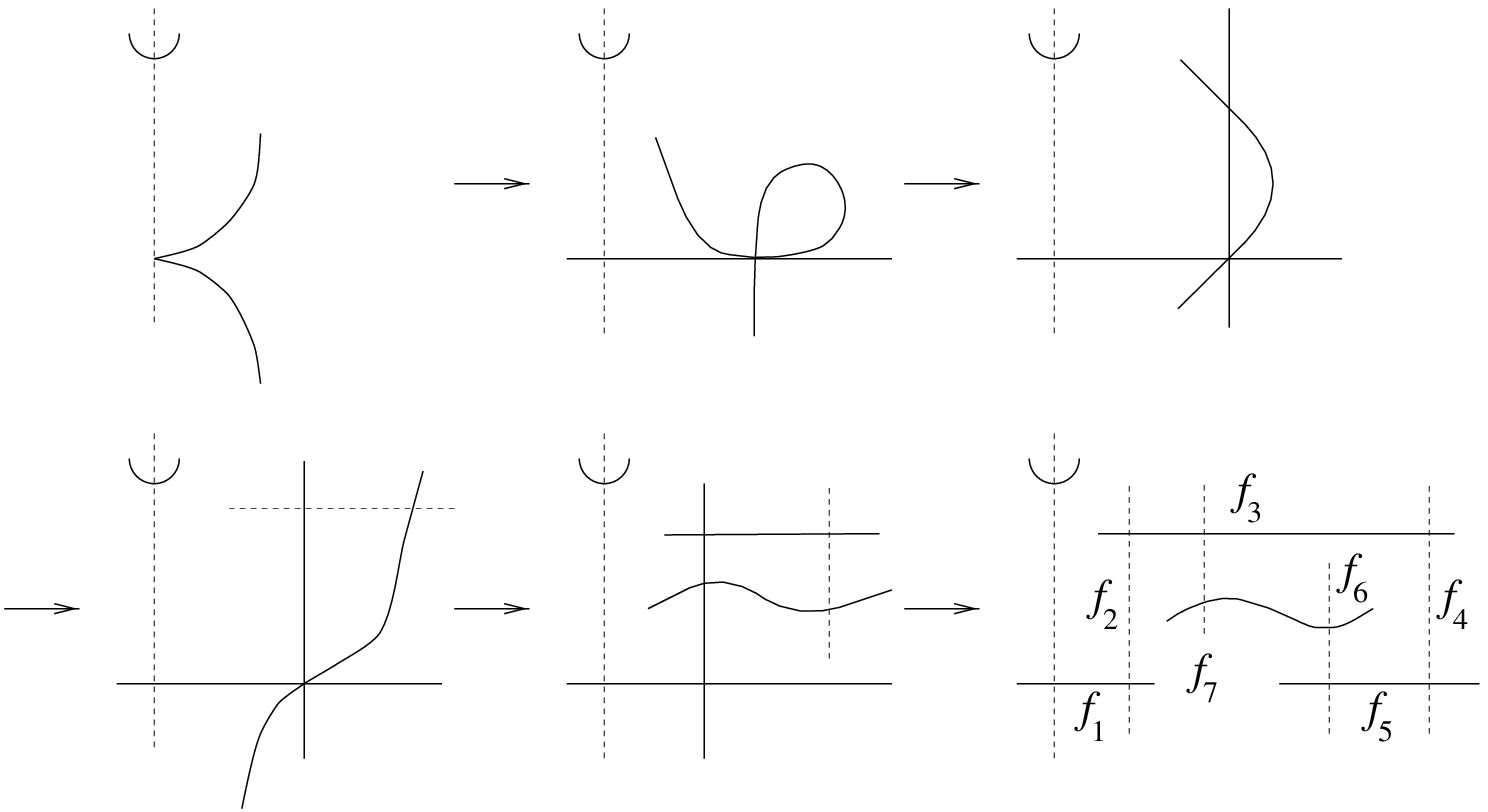}}

\ifig\exM{Partial resolution of the $II^*$ fiber over the collision 
locus. The limitations of two dimensional drawing prevent a correct 
representation of the original singular branch locus.}
{\epsfxsize3.7in\epsfbox{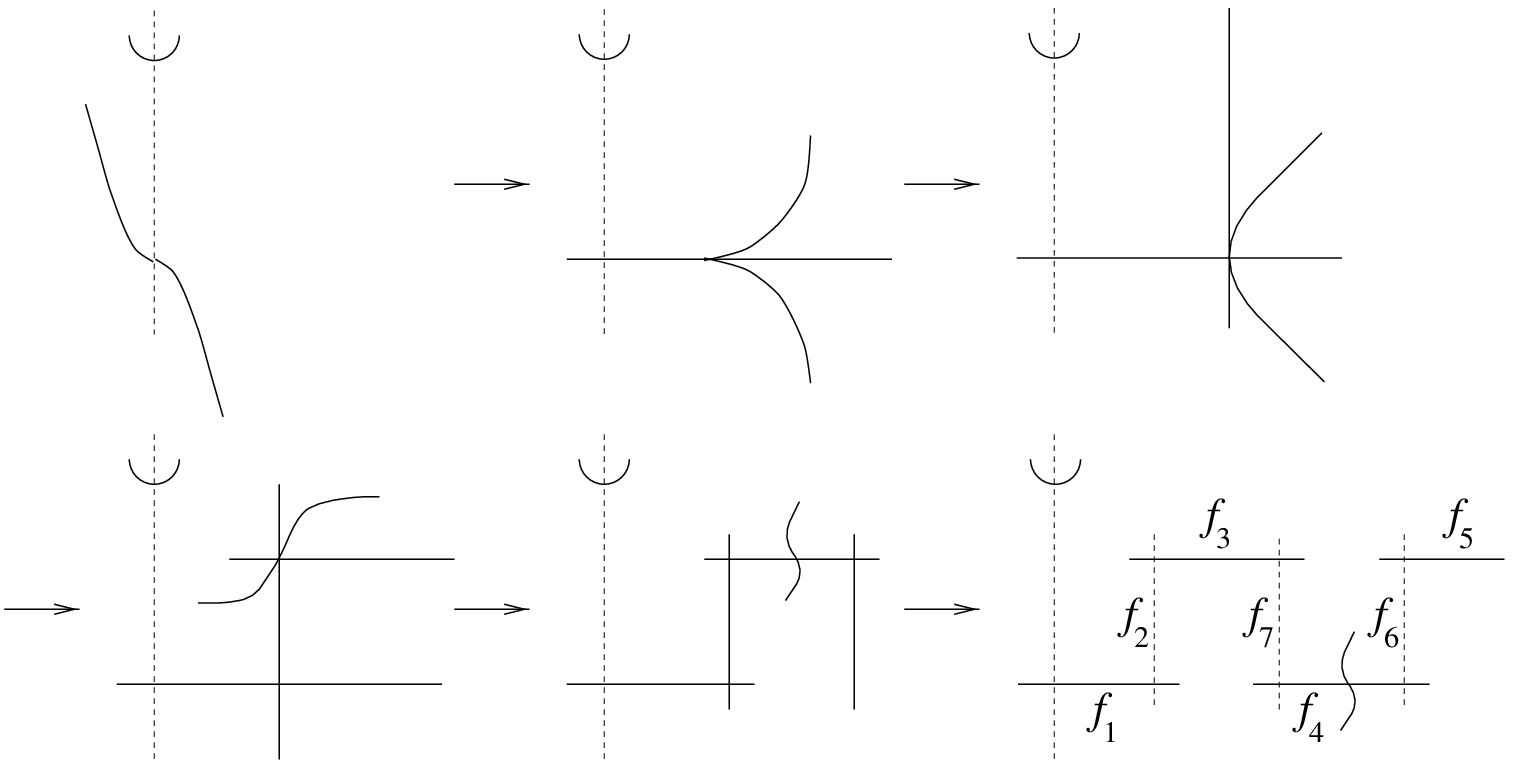}}

The sequence of blow-ups 
leads to a partial resolution of the $II^*$ fiber over the 
collision locus as in fig. 20. Note that the branch locus
intersects itself over the collision locus. However, the branch
surface is of analytic type $uv=w$ near the intersection point. 
As this is smooth, the double cover is also smooth and no further
blow-ups are necessary. 
Taking into account the structure of the 
singular fiber over the collision point, it follows that this is
precisely the first $E_7$ degeneration introduced in section 4.

\listrefs
\end